\documentclass[floatfix,twocolumn,amsmath,amssymb,aps,prb,showpacs,superscriptaddress]{revtex4-1}
\usepackage{amsmath}
\usepackage{graphicx}
\usepackage{times}
\usepackage{subfigure}
\usepackage{color}
\usepackage[sort&compress]{natbib}

\newcommand{\bs}[1]{\boldsymbol{#1}}

\newcommand{\mth}{\mathcal{H}}
\newcommand{\mss}{\mathsf{S}}

\newcommand{\mhz}{\mathbf{\hat{z}}}
\newcommand{\mhx}{\mathbf{\hat{x}}}

\newcommand{\eto}{Er$_{2}$Ti$_{2}$O$_{7}$}

\begin{document}

\title{Order Induced by Dilution in Pyrochlore XY Antiferromagnets}
\author{A. Andreanov}
\affiliation{Max-Planck-Institut f\"{u}r Physik komplexer Systeme, D-01187 Dresden, Germany}
\author{P. A. McClarty}
\affiliation{ISIS Neutron and Muon Source, STFC, Rutherford-Appleton Laboratory, Harwell Campus, Oxfordshire, OX11 0QX, UK}

\date{\today}

\begin{abstract}
	XY pyrochlore antiferromagnets are well-known to exhibit order-by-disorder through both quantum and thermal selection. In this paper we consider the effect of substituting non-magnetic ions onto the magnetic sites in a pyrochlore XY model with generally anisotropic exchange tuned by a single parameter $J^{\pm\pm}/J^\pm$. The physics is controlled by two points in this space of parameters $J^{\pm\pm}/J^\pm=\pm 2$ at which there are line modes in the ground state and hence an $O(L^2)$ ground state degeneracy intermediate between that of a conventional magnet and a Coulomb phase. At each of these points, single vacancies seed pairs of line defects. Two line defects carrying incompatible spin configurations from different vacancies can cross leading to an effective one-dimensional description of the resulting spin texture. In the thermodynamic limit at finite density, we find that dilution selects a state``opposite" to the state selected by thermal and quantum disorder which is understood from the single vacancy limit. The latter finding hints at the possibility that Er$_{2-x}$Y$_x$Ti$_2$O$_7$ for small $x$ exhibits a second phase transition within the thermally selected $\psi_2$ state into a $\psi_3$ state selected by the quenched disorder.
\end{abstract}

\pacs{
	74.62.Dh
	75.10.Hk
	75.50.Ee
}

\maketitle

\section{Introduction}

Heisenberg antiferromagnets on lattices of corner-sharing triangles such as the kagome, hyperkagome and pyrochlore lattices with exchange couplings between nearest-neighbors that, in certain instances, lead to a classical ground state degeneracy that scales with the linear system size $L$ as $O(L^{d\alpha})$ with $0<\alpha\leq 1$ where $d$ is the dimensionality of the model.\cite{lacroix2011introduction} This is the core principle of \textit{geometrically frustrated} magnetism which stands in contrast to conventional unfrustrated magnetism in which there are $O(1)$ classical ground states. In practice, the effect of geometrically frustrated interactions is to push down the leading magnetic energy scale and impose a local constraint on the magnetism that, in a departure from trivial paramagnetism, implies strong correlations between the moments. 

One of the characteristics of geometrically frustrated magnets is that over some temperature interval below the Curie-Weiss temperature, the local constraint is important in determining the phenomenology.\cite{lacroix2011introduction} For example, in spin ices, a local divergence-free condition leads to characteristic ``pinch-point" correlations and excitations that interact through a Coulomb interaction of entropic origin.\cite{lacroix2011introduction,henleycoulomb2010,castelnovo2012spinice} The classical degeneracy is typically broken by sub-leading interactions generating potentially many competing instabilities including those into novel states of quantum matter.\cite{balents2010spinliquid} It is also possible for a sub-extensive classical ground state degeneracy to be lifted by fluctuations leading to discrete symmetry breaking at finite temperature. This mechanism, which is reminiscent of anomalies that arise in certain quantum field theories\cite{bertlmann2000anomalies}, is known as \textit{order-by-disorder}.\cite{villain1980order,henley1989ordering,prakash1990ordering,sheng1992ordering,bergman2007order}

Order-by-disorder is a feature of a number of models of frustrated magnets.\cite{villain1980order,henley1989ordering,prakash1990ordering,sheng1992ordering,bergman2007order} In isotropic Heisenberg models, the expectation\cite{henley1989ordering} is for order-by-disorder to select the most collinear magnetic structure. The intuition behind this is that fluctuations will couple moments most strongly if they are collinear.\cite{henley1989ordering} The fragility of the classical ground state degeneracy to fluctuations extends also to dilution which can also induce long-range order. In many cases where it has been investigated, order-from-lattice-disorder occurs into a discrete set of states that are ``opposite" to those selected by fluctuations.\cite{henley1989ordering,prakash1990ordering} 

Rare-earth pyrochlore magnets exhibit much of the richness that one would expect of geometrically frustrated magnets on the general grounds mentioned above.\cite{gardner2010magnetic} Over recent years, we have understood that the correct framework within which to understand these systems is a model with anisotropic exchange couplings between nearest-neighbors perhaps supplemented with further neighbor couplings or the dipolar interaction.\cite{curnoestructural2008,mcclarty2009energetic,thompson2011rods,ross2011quantum,zhitomirsky2012quantum,savary2012order} It is known that the nearest-neighbor model exhibits order-by-disorder induced by both quantum and thermal fluctuations over a significant range of the available space of couplings.\cite{wong2013ground,yan2013living} The fact that this phenomenon occurs over a wide region of the parameter space in the model hints that it might also occur in real materials and indeed, in \eto, the low temperature magnetic structure\cite{champion2003er,poole2007magnetic} can arise only through the effect of fluctuations.\cite{mcclarty2009energetic,zhitomirsky2012quantum,savary2012order,petit2014orderbydisorder} In this case of order-by-disorder, there is no sense in which the selected state is the most collinear. Indeed, the usual effective biquadratic term $\sum_{\langle ij\rangle} (\mathbf{S}_i \cdot \mathbf{S}_j)^2$ generated by fluctuations is a constant within the set of classical XY ground states of \eto.\cite{mcclarty2014order} 

Order-by-disorder in \eto\ is controlled by a point in the space of possible nearest-neighbor XY couplings at which the ground state degeneracy scales as $O(L^2)$.\cite{mcclarty2014order,wong2013ground,yan2013living} At this point, there are four continuous global $U(1)$ symmetries in the classical ground state which intersect at discrete configurations.  Fluctuations stabilize the magnetic structure precisely at those intersection points.\cite{mcclarty2014order,wong2013ground,yan2013living}

In this paper, we consider the effect of dilution in the most general nearest-neighbor XY antiferromagnet on a pyrochlore lattice parametrized by some coupling $\eta\equiv J^{\pm\pm}/J^\pm$. This model has a pair of high symmetry points at $\eta=\pm 2$ including the one alluded to above. We study first the effect of putting a single vacancy into a finite system. At the high symmetry points, point-like impurities leads to line-like defects extending through the system that become localized away from $\eta=\pm2$. In any case, the continuous symmetries are broken down to a discrete set of states ``opposite" to those selected by thermal fluctuations. We study the approach to the thermodynamic limit, showing that this situation persists in the thermodynamic limit for a finite density of vacancies.

The next section is an introduction to the model, its classical ground states and the mechanism of order-by-disorder in this model. Our goal is to study the thermodynamic limit in the case of a finite density of non-magnetic impurities on the pyrochlore sites. We approach this limit by considering, in Section~\ref{sec:VandT}, the problem of removing spins from a single tetrahedron or from a pair of tetrahedra at zero and finite temperatures where we uncover phenomena intermediate between conventional non-collinear magnets and Coulomb phases. Then, in Section~\ref{sec:finite_density}, we consider the problem of a finite density of vacancies before offering our conclusions (Section~\ref{sec:conclusions}). 

\section{Model}

We consider the class of pyrochlore magnets with strong spin-orbit coupling and where the low-lying crystal field states form a well separated doublet. Many of the rare-earth pyrochlore magnets fall within this class of states. The leading order phenomenology of these magnets can be described on the basis of the most general nearest-neighbor exchange interactions allowed by symmetry.\cite{curnoestructural2008,mcclarty2009energetic,thompson2011rods,ross2011quantum,savary2012order} Following the convention of Ref.~\onlinecite{ross2011quantum}, this model is
\begin{align}
	\mth = & \sum_{\langle ij\rangle} J^\mathrm{zz} \mss^z_i  \, \mss^z_j + J^\pm \left(\mss^+_i  \, \mss^-_j + \textrm{h.c.}\right) \notag\\
	& + J^{\pm\pm} \left( \gamma_{ij} \mss^+_i  \, \mss^+_j  + \textrm{h.c.}\right)\notag\\
	& -  J^{z\pm} \left\{ \mss^z_i  \left( \zeta_{ij} \mss^+_j  + \, \zeta^{*}_{ij} \mss^-_j\right) + \left( i \leftrightarrow j \right) \right\} 
	\label{eqn:H}
\end{align}
where the components are taken within local coordinate frames - one for each member of the tetrahedral basis - with the $z$ axes along the $[111]$ directions. The matrices $\gamma_{ij}$ and $\zeta_{ij}$ are unimodular complex $4\times 4$ matrices. Both the local coordinate frame and interaction matrices are given in Appendix~\ref{sec:appendix}. Stringent tests based on extensive inelastic neutron scattering data in an ordered phase and bulk measurements have found excellent agreement between the predictions of this model and experiment in the cases of Yb$_2$Ti$_2$O$_7$\cite{ross2011quantum,applegate2012vindication,hayre2013thermodynamic} and Er$_2$Ti$_2$O$_7$.\cite{savary2012order,zhitomirsky2012quantum,oitmaa2013phase} Moreover, this model exhibits ground states in zero field which have been found in a number of other pyrochlore magnets for which the exchange parameters have not yet been determined.\cite{yan2013living}

In the following, we restrict our attention to a slice through the parameter space of this model. In particular we take $J^\mathrm{zz}=J^{z\pm}=0$ so that we are left with the most general XY model in which the easy planes, which are perpendicular to the local Ising axes, are non-coplanar. This model has a single tuning parameter $J^{\pm\pm}/J^{\pm}$ up to an overall scale. This is the simplest model that exhibits the physics that we wish to explore. Perfect XY models (those with no Ising component of the exchange) at the single ion level are fine tuned because there is no symmetry to force $\hat{J}^z$ matrix elements within the single ion doublet to be zero. Nevertheless, the physics we discuss in this paper survives the presence of sufficiently weak $J^\mathrm{zz}$ and $J^{z\pm}$ perturbations because it is dependent on the presence of soft modes which will be introduced in the next subsection. The ``softness" of these modes varies smoothly in parameter space for a given set of ground states which themselves survive such perturbations.

For the XY model, we parametrize spins $\mss_i$ by their angle $\phi_i$: $\mss_i = (\cos\phi_i,\sin\phi_i,0)$ in the local frame.

\begin{figure*}[htpb]
	\includegraphics[width=1.7\columnwidth]{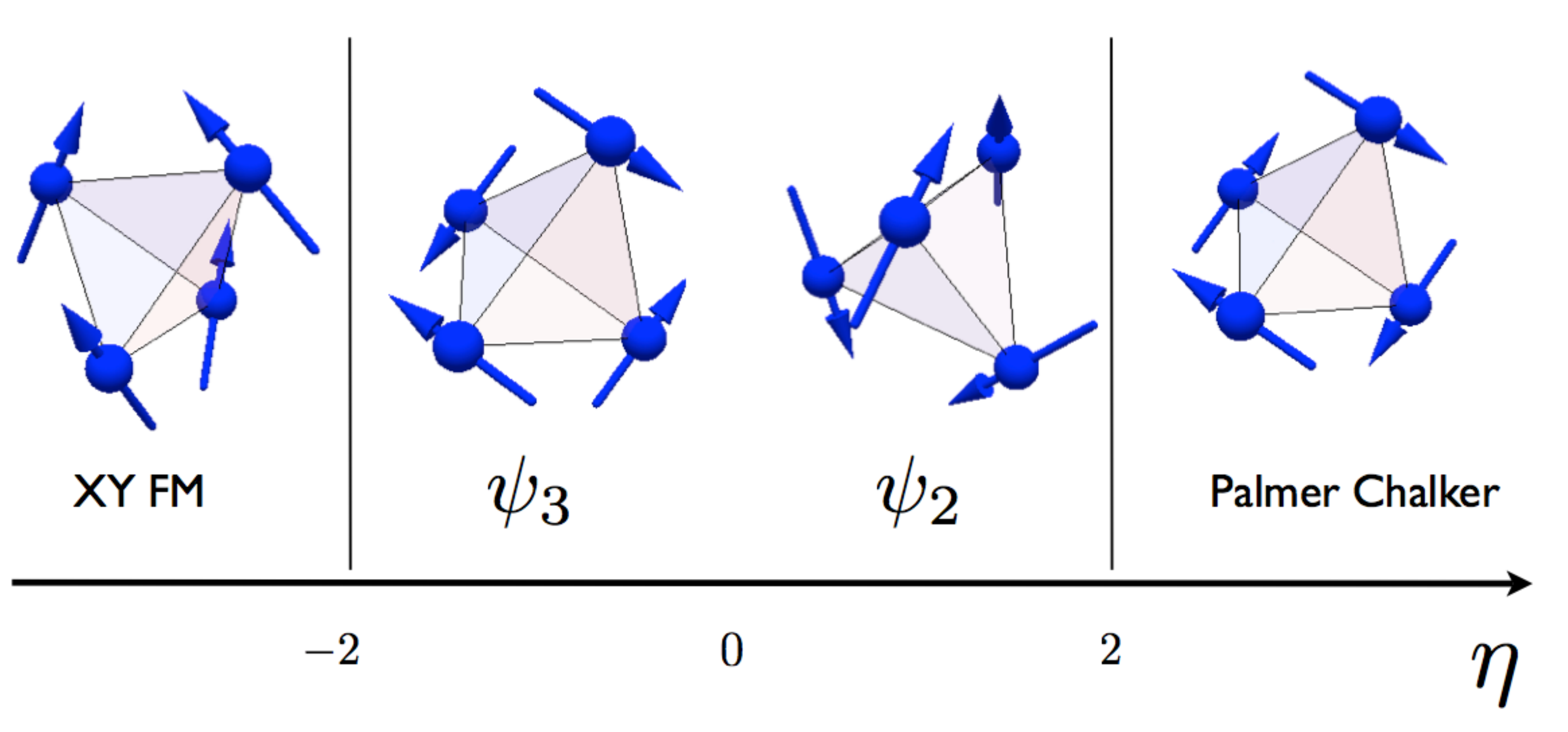}
	\caption{
	\textbf{ Phase diagram of generalized XY model.} (Color online). The phase diagram for the generalized XY model as a function of tuning parameter $\eta$ defined in the text, that interpolates between two highly degenerate points $\eta=\pm2$. Thermal fluctuations select the $\psi_2$ structure for $0<\eta\leq 2$ and $\psi_3$ for $-2\leq\eta< 0$. The magnetic structures are indicated. The point $\eta=0$ is an isotropic XY magnet in the local coordinate frame and the order-by-disorder selection mechanism gives way to ordinary global $U(1)$ spontaneous symmetry breaking at that special point.
	}
	\label{fig:XYModel}
\end{figure*}

\begin{figure}[htpb]
	\includegraphics[width=\columnwidth,angle=90]{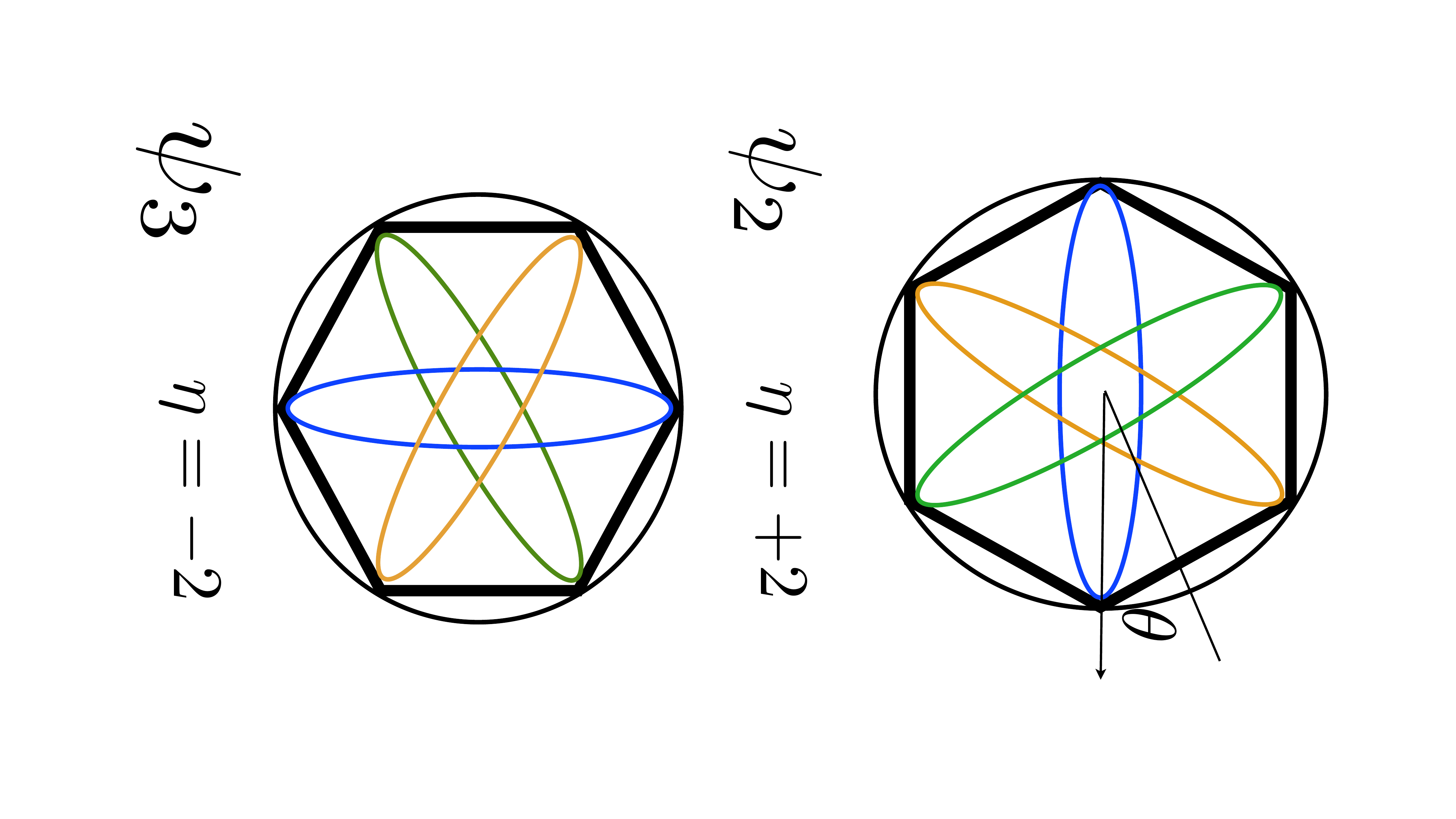}
	\caption{
	\textbf{Degeneracies at $\eta=\pm 2$} (Color online). The plot represents the continuous degeneracies present in the model and given explicitly in Eqs.~\eqref{eq:eta2} and~\eqref{eq:etam2}. There is one principal branch of continuous degeneracies (the outer ring) with pairs of discrete intersection points with each of three other $U(1)$ degeneracies which are denoted by colored ovals within the main hexagon. The intersection points (at the hexagonal vertices) determine the thermal selection - either $\psi_2$ or $\psi_3$. The ground states at couplings $\eta=2$ and $\eta=-2$ are related by a simple transformation given in the main text and the branches represented with identical colors in this figure are related by this mapping.   
	}
	\label{fig:XYModel2}
\end{figure}

\subsection{Special Points and their Ground States}
\label{sec:branches}

It turns out that the physics over most of the phase diagram of the XY model, which is depicted in Fig.~\ref{fig:XYModel}, is actually controlled by a pair of points in the space of couplings. The first of these is obtained by taking the Heisenberg model with isotropic exchange $J$ and projecting the spins onto the local XY planes. In terms of the couplings in Eq.~\eqref{eqn:H}, this point corresponds to $J^\pm=-J/6$ and $J^{\pm\pm}=J/3$. The second special point we consider is at $J^\pm=-J/6$ and $J^{\pm\pm}=-J/3$. In terms of variable $\eta\equiv \vert J^{\pm\pm} \vert/J^\pm$ the special points are at $+2$ and $-2$ - the former corresponding to the projected Heisenberg model. The ground states for $\eta=2$ are given in Ref.~\onlinecite{mcclarty2014order}. In the local coordinate convention of this paper, the ground states on a single tetrahedron with local XY angles $ \left( \phi_1,\phi_2,\phi_3,\phi_4 \right)$, which can be found by requiring that the three components of the total moment in the laboratory frame equal zero, 
\begin{align}
	\label{eqn:GS1}
	B1: & \left( \theta,\theta,\theta,\theta \right)  \\
	B2: & \left( \theta,\theta,-\theta,-\theta \right)  \notag \\
	B3: & \left( \theta,2\pi/3 - \theta,2\pi/3 - \theta,\theta \right)  \notag \\
	B4: & \left( \theta,4\pi/3- \theta,\theta,4\pi/3 - \theta \right) \notag
	\label{eq:eta2}
\end{align}
where $\theta\in [-\pi,\pi)$ is measured relative to the local $\bs{\hat{x}}$ axis on a given sublattice. There are four global symmetries in this model at the classical level because we can tile the lattice with identical tetrahedra in one of these four branches and rotate the spins smoothly within a branch remaining in the ground state. Apart from these global $U(1)$ symmetries, there are line modes. To see this, suppose the spins are arranged in the equal angle branch for some $\theta$. Then we can choose a chain running through the lattice. The sublattice labels alternate between two values on the chain and we can transform the tetrahedra along this chain into a new branch. For example, if the chain runs along the $2,3$ sublattices, we can take these sublattices from branch one to branch two with $\theta\rightarrow -\theta$ and remain in the ground state. It follows that the number of ground state degrees of freedom scales with $L^2$ in contrast to both conventional magnets, which have $O(1)$ ground states, and 3D models exhibiting Coulomb phases which have an $O(L^3)$ ground state entropy.

Consider now $\eta=-2$. At this coupling, the Hamiltonian on a single tetrahedron in the lab frame is identical to that for the $\eta=2$ point after a transformation taking the angle in the XY plane $\phi_i \rightarrow \phi_i + \pi/2$ for all four sublattices. This mapping is represented in Fig.~\ref{fig:XYModel2}. It follows that there are four branches of ground states with
\begin{align}
	B1: & \left( \theta,\theta,\theta,\theta \right)  \\
	B2': & \left( \theta,\theta ,\pi-\theta,\pi-\theta \right) \notag \\
	B3': & \left( \theta,5\pi/3 - \theta,5\pi/3 - \theta, \theta \right) \notag \\
	B4': & \left( \theta,\pi/3- \theta,\theta,\pi/3 - \theta \right) \notag
	 \label{eq:etam2}
\end{align}
which also have associated line modes. The branch $B1$ is antiferromagnetic - the net moment is zero on the tetrahedron. For the remaining three branches at $\eta=-2$, the tetrahedron has a net moment except for special values of $\theta$.

\subsection{Ground States of the Generalized XY Model}

The equal angle branch (number one of Eq.~\eqref{eqn:GS1}), $B1$, which lies in the ground state at the points $\eta=\pm 2$ is, at the classical level, robust to all perturbations of the form $J_{ij}^{\alpha\beta}\mss_i^\alpha\mss^\beta_j$ which respect the lattice symmetries as shown for the nearest-neighbor case in Ref.~\onlinecite{mcclarty2009energetic} and generalized in Ref.~\onlinecite{savary2012order}. This means that degenerate branch $B1$ of Eq.~\eqref{eq:eta2} lies somewhere in the classical spectrum. For the range $-2<\eta<+2$, the $B1$ branch is the only set of ground states - the other branches are gapped out away from $\eta=\pm 2$ (see panel (a) of Fig.~\ref{fig:OBD}). For $\eta > 2$, the ground states are the Palmer-Chalker states\cite{palmer2000order} and for $\eta < -2$, the ground state configuration is a ferromagnet with spins in the XY plane.\cite{wong2013ground} The absence of other branches besides the equal angle branch in the interval $-2<\eta<+2$,  implies that there are no line modes among the ground states. However, the ground states and low-lying modes for couplings $\eta$ and $-\eta$ are related by a $\pi/2$ rotation in the local XY angles so our conclusions for $0<\eta<2$ have a straightforward analog for $-2<\eta<0$.

\subsection{Order by Disorder}
\label{sec:ObD}

\begin{figure}
	\includegraphics[width=0.95\columnwidth]{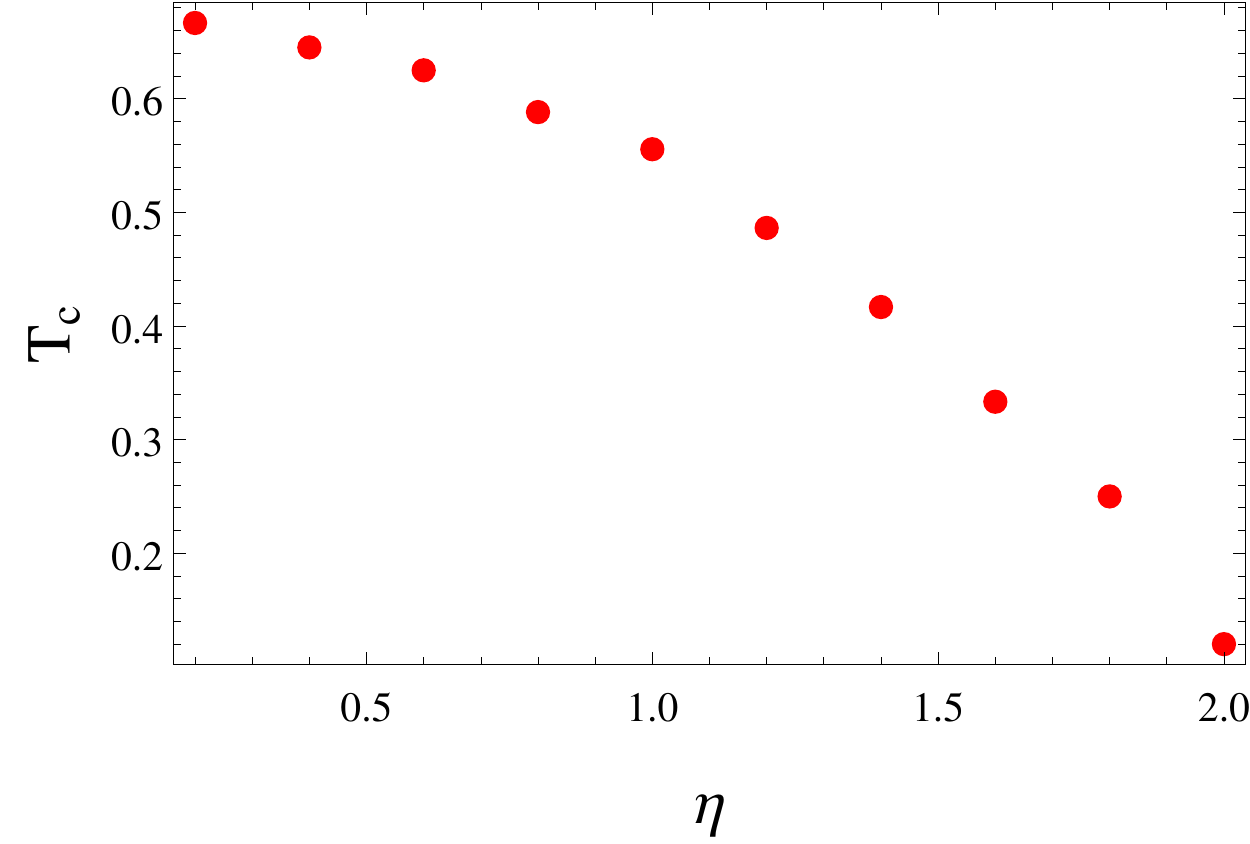}
	\caption{
	\textbf{Sketch of $T_c$ vs $\eta$ for the clean system}. (Color online). The $T_c$ corresponds to a transition from the paramagnet into the $\psi_2$ state. These data were taken from a Monte Carlo simulation with $432$ spins with the transition temperature drawn from the location of the heat capacity peak. At the isotropic point $\eta=0$, the finite temperature transition proceeds via ordinary spontaneous symmetry breaking and not by order-by-disorder.
	}
	\label{fig:TcSketch}
\end{figure}

The classical degeneracies of the generalized XY model are resolved by thermal fluctuations leading to long-range magnetic order below some $T_c$.\cite{champion2003er,champion2004soft,mcclarty2014order,zhitomirsky2012quantum,yan2013living} The phase diagram together with representations of the ordered magnetic structures are shown in Fig.~\ref{fig:XYModel}. In the figure the $\psi_2$ and $\psi_3$ states are the states selected by order-by-disorder. Both of these states have spin configurations in the equal angle branch $B1$ of Eq.~\eqref{eq:eta2}. In the pyrochlore XY model, the selected states are $\psi_2$ for $0<\eta\leq 2$ and $\psi_3$ for $-2\leq \eta<0$ (see Fig.~\ref{fig:XYModel}). The emergence of these ordered structures, which cannot be collinear in contrast to examples of order-by-disorder in Heisenberg magnets, can be understood intuitively from the ground states at the two special couplings $\eta=\pm 2$. For example, at $\eta=+2$, one observes that branch 1 intersects with another branch at exactly two angles giving six intersections in total - at $\theta=n\pi/3$ for $n=0\ldots 5$ - which correspond to the six discrete domains of $\psi_2$. Thermal fluctuations select the softest modes which are at the intersection points. This is exactly the result obtained from classical Monte Carlo simulation of the $\eta=+2$ model\cite{champion2003er,champion2004soft,mcclarty2014order} and deformations about it.\cite{zhitomirsky2012quantum,yan2013living} The phase diagram of the $0<\eta<2$ model is shown in Fig.~\ref{fig:TcSketch}. Similarly, at $\eta=-2$, the intersection points at $\theta=\pi/6 + n\pi/3$ belong to the thermally selected $\psi_3$ states. 

\begin{figure}[htpb]
\begin{subfigure}[][]{
\includegraphics[width=0.95\columnwidth]{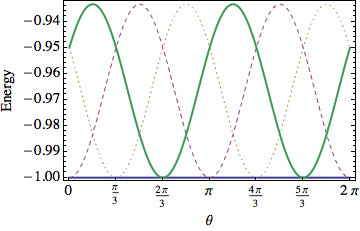}
}
\end{subfigure}
\begin{subfigure}[][]{
\includegraphics[width=0.95\columnwidth]{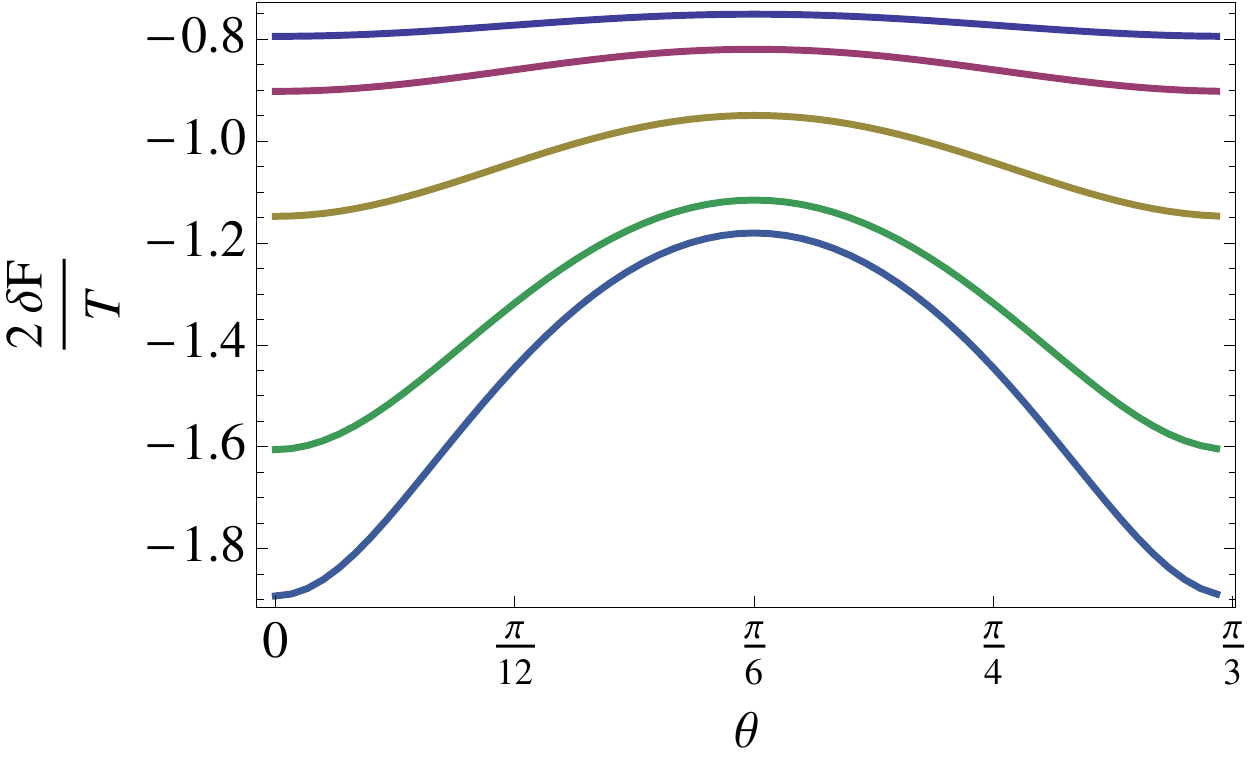}
}
\end{subfigure}
	\caption{
	\textbf{Ground states and thermal selection in the generalized XY model.} (Color online). 
	(a) Plot of the energies of the four branches of Eq.~\eqref{eqn:GS1} for $\eta=1.9$ showing that the degeneracies of three branches has been lifted weakly although the touching points at the $\psi_2$ configurations $n\pi/3$ remain. The curves are for branch $B2$ (dashed), $B3$ (dotted), $B4$ (solid) and branch $B1$ is the horizontal line of lowest energy. 
	(b) Plot of $(2/T) \delta F$ for the equal angle branch $B1$ of Eq.~\eqref{eqn:GS1} for various $\eta$ in the range $0$ to $2$ ($\eta$ increases from top to bottom) showing that fluctuations break the degeneracy in favor of the $\psi_2$ states. 
	}
	\label{fig:OBD}
\end{figure}

We can also understand the thermal selection on the basis of a perturbative calculation.\cite{mcclarty2014order} We expand Eq.~\eqref{eqn:H} in powers of small angular fluctuations and compute the reduced free energy about some classical ground state. Thermal selection is determined by the spectrum of the Hessian matrix $A^{ab}_{\mathbf{q}}$. The reduced free energy relative to the classical ground state energy, within this low temperature expansion, is\cite{mcclarty2014order}
\begin{gather}
	\delta F = \frac{T}{2} \sum_{\mathbf{q}, a,b} \log \left[ \mathrm{Det} \mathbf{A}_\mathbf{q} \right].
\end{gather}
The sum over reciprocal space is illustrated for different branch one (Eq.~\eqref{eqn:GS1}) ground states for different values of $\eta$ in Fig.~\ref{fig:OBD}(b). The minima for $0< \eta<2$ are at the $\psi_2$ angles $\theta=n\pi/3$. At the $\psi_2$ ground state for $\eta=+2$, the line modes in real space appear as planes of zero modes in reciprocal space which determine the selection. The line modes are soft modes for $\eta$ away from the high degeneracy point $\eta=+2$.

Quantum fluctuations have been studied about the classical ground states. One finds that the leading order correction to the ground state energy coming from fluctuations selects the same state as thermal fluctuations.\cite{mcclarty2014order,zhitomirsky2012quantum,savary2012order,wong2013ground,yan2013living} The mechanism is analogous to the thermal selection mechanism: the softest modes lie at the $\psi_2$ configurations for $0<\eta\leq 2$ and hence the contribution to the zero point energy is least.

\section{Vacancies and Textures}
\label{sec:VandT}

In this section, we begin by considering the effect of removing one, two and three spins from a single tetrahedron in the XY pyrochlore antiferromagnet firstly at $\eta=\pm 2$ and then away from these highly degenerate points. In the second subsection, we consider the simplest spin texture at these highly degenerate points.

The effect of creating vacancies on the background spin configuration in magnetic systems has been studied for many years (see for example, Ref.~\onlinecite{villain1979insulating,sen2011fractional,wollny2011fractional}). The symmetric points $\eta=\pm 2$ in our model demonstrate a novel feature: vacancies break down the continuous degeneracies of the background and introduce line defects into the system preserving zero moment on all tetrahedra. In this case, the approach to the thermodynamic limit with a finite density of defects is far from obvious because the line defects naively become dense long before the thermodynamic limit is reached. In order to gain some insight about the nature of the ground states in the thermodynamic limit, we consider the simplest non-trivial spin texture where line defects of incompatible angles intersect. 

Moving away from the symmetric points, the line defects are penalized energetically and acquire a length scale. In this case, the single vacancy problem considered below gives us direct insight into the nature of the finite density problem in the thermodynamic limit. 

\subsection{Vacancies}

Let us consider the coupling $\eta=+2$ at which there is an $O(L^2)$ ground state entropy and remove a single spin from a finite periodic system. The zero moment conditions on a single tetrahedron without any vacancy are
\begin{align}
	\cos\theta_1 + \cos\theta_2 - \cos\theta_3 - \cos\theta_4 = 0 \notag \\
	\cos \bar\theta_1  - \cos \bar\theta_2  + \cos\bar\theta_3 - \cos\bar\theta_4  = 0 \notag \\
	\cos\hat\theta_1  - \cos \hat\theta_2   - \cos\hat\theta_3   + \cos \hat\theta_4  = 0
	\label{eqn:ZeroMoment}
\end{align} 
where $\bar\theta=\theta+(\pi/3)$, $\hat\theta=\theta+(2\pi/3)$ and the subscripts refer to different sublattices. These can be solved to obtain the four branches given in Eq.~\eqref{eq:eta2}. When we remove a spin, the equations can be solved for the remaining three spins to give a discrete set of solutions. There are eight such solutions for a spin missing on a given sublattice or four solutions modulo time reversal. The angles that appear in the solutions are all of the form $\pi/6 + n\pi/3$. For example, if we remove site $1$, one of the solutions preserving zero moment is $\left(-,\frac{\pi}{6},-\frac{\pi}{2},-\frac{\pi}{6}\right)$. In order to determine the possible ground states on a lattice, we have to find solutions that are compatible with the branches of Eq.~\eqref{eqn:GS1}. In order to construct a zero moment solution, we may choose some background angle which the single tetrahedron solutions constrain to be of the form $\pi/6+ n\pi/3$. Then, of the eight solutions to Eq.~\eqref{eqn:ZeroMoment}, only two are compatible with the ground states on the lattice for a given choice of background angles. For example, if we choose a $\theta=-\pi/6$ background with all angles equal (i.e. lying within branch one of Eq.~\eqref{eqn:GS1}), zero moment overall is preserved by having a pair of line defects radiating from the vacant site with tetrahedra in the configuration $(\pi/6,\pi/6,-\pi/6,-\pi/6)$ and $(-\pi/2,-\pi/6,-\pi/2,-\pi/6)$ (again, angles correspond to sublattices $1,2,3,4$ respectively). The background $\psi_3$ configuration retains a discrete degeneracy: as discussed in Section~\ref{sec:branches}, one may include line modes that extend across the system and that do not cross. Further details are provided in Appendix~\ref{sec:appendix2}.

\begin{figure}[htpb]
	\includegraphics[width=\columnwidth,angle=-90]{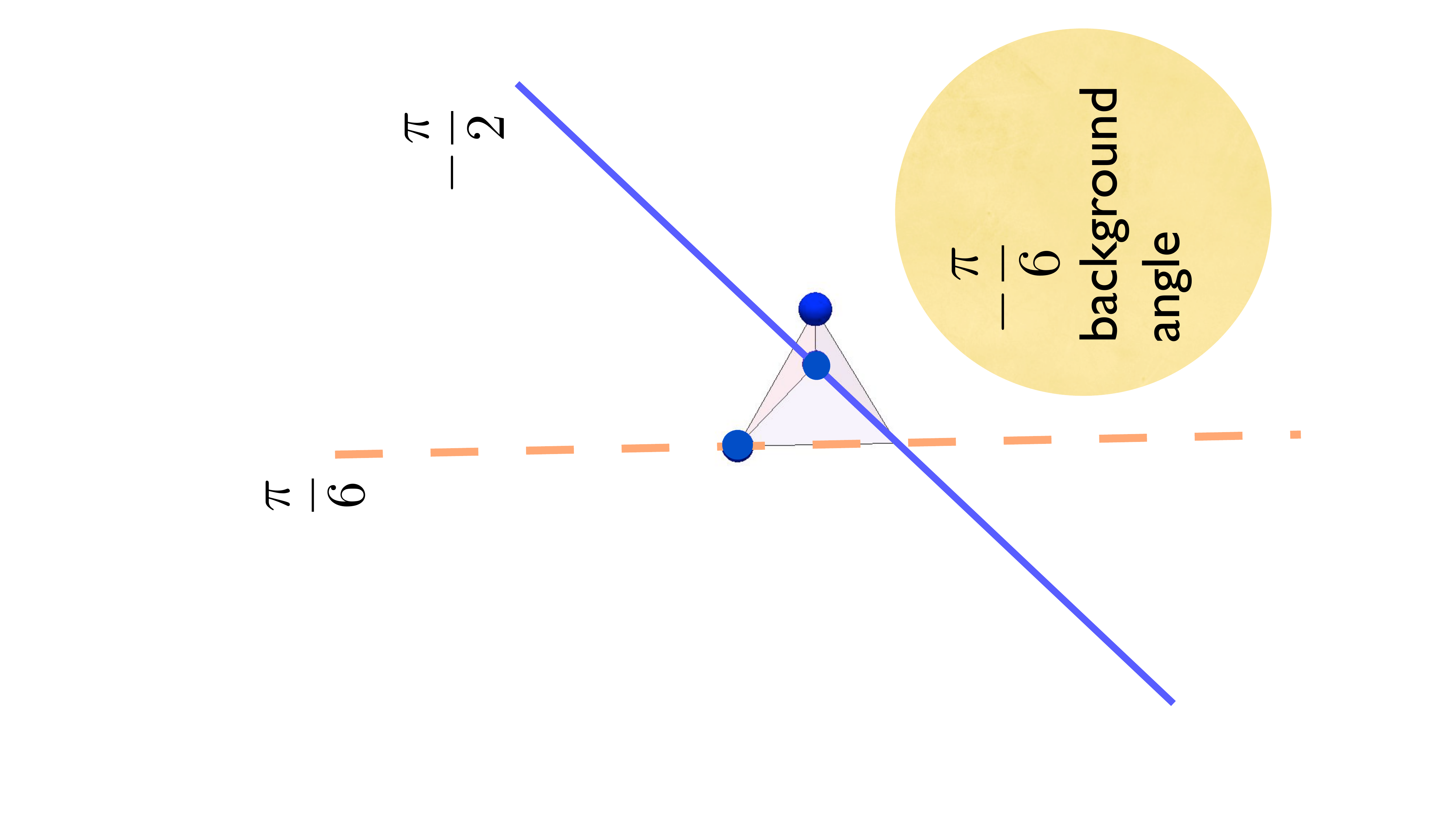}
	\caption{
	\textbf{Schematic plot of a single vacancy spin ground state configuration at $\eta=2$} (Color online). The figure shows a single tetrahedron in the pyrochlore lattice with a single spin missing. There are spins on the blue spots on three of the vertices. All other magnetic sites have been suppressed to avoid cluttering the figure. The presence of the vacancy selects a background angle among the set of $\psi_3$ states in the local coordinate frame of the spins. Here this is set to $\theta=-\pi/6$ in the equal angle branch $B1$ of Eq.~\eqref{eq:eta2} which is shared by all spins away from the line defects (blue and orange lines). The angles refer to the local XY coordinate frame given in Appendix~\ref{sec:appendix}. The vacancy produces two line defects passing through the vacancy along chains in the lattice - one chain with spins lying at angle $\pi/6$ (orange,dashed) and one with spins at angle $-\pi/2$ (blue). This means that two chains of tetrahedra have spins with configurations in one of the branches $B2-4$. 
	}
	\label{fig:Vacancy}
\end{figure}

In summary, the presence of a single vacancy in a finite system for $\eta=+2$ breaks down the continuous degeneracies completely leading to a uniform $\psi_3$ background (up to line modes avoiding the vacancy) and two line defects passing through the vacancy along which the tetrahedra have configurations belonging to one of the three non-uniform angle branches in Eq.~\eqref{eqn:GS1}. This configuration is represented in Fig.~\ref{fig:Vacancy}. All tetrahedra in the system including the two sharing the vacancy have zero net moment in the laboratory frame. Provided line defects do not intersect, further sites may be removed and even placed directly on line defects while retaining zero moment overall. In the following section, we pick up on these insights to consider the finite density case. 

As we have noted in Section.~\ref{sec:branches}, one simply rotates all the discrete angles above by $\pi/2$ to find the ground state configurations at $\eta=-2$. The only difference between $\eta=2$ and $\eta=-2$ is that the tetrahedra along line defects for $\eta=-2$ have nonzero net moment. Bearing these points in mind, the discussion above carries over straightforwardly to the $\eta=-2$ coupling.

\begin{figure}[htpb]
	\includegraphics[width=0.75\columnwidth]{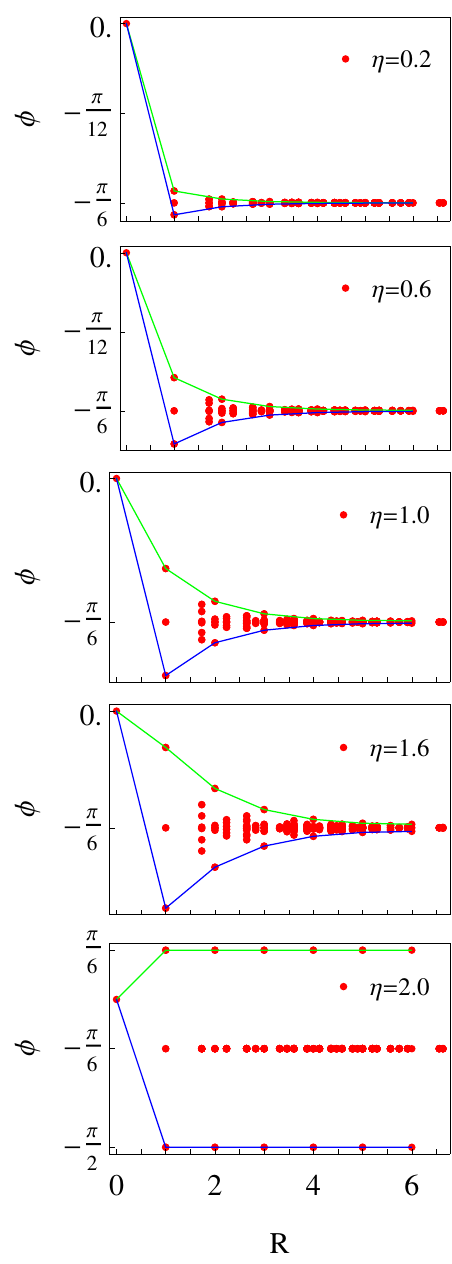}
	\caption{
	\textbf{Series of panels showing local angles of all the spins with varying distance from the vacancy.} (Color online). The parameter $\eta$ increases from top to bottom. $\eta=2$ is at the bottom and the plot shows that the single vacancy texture is strongly anisotropic with two line defects running through the vacancy. The angles corresponding to line defects passing through the vacancy are joined with solid lines. The background (far from the line defects) has a discrete degeneracy and the figures correspond to a common choice of the background angle ($\theta=-\pi/6$). For $0<\eta\neq 2$, the vacancy texture acquires a length scale that decreases as $\eta$ approaches $0$. It also becomes more isotropic as $\eta$ approaches $0$. The angle at the vacancy has been arbitrarily set to zero in these figures.
	}
	\label{fig:SV}
\end{figure}

\begin{figure}[htbp]
	\includegraphics[width=0.95\columnwidth]{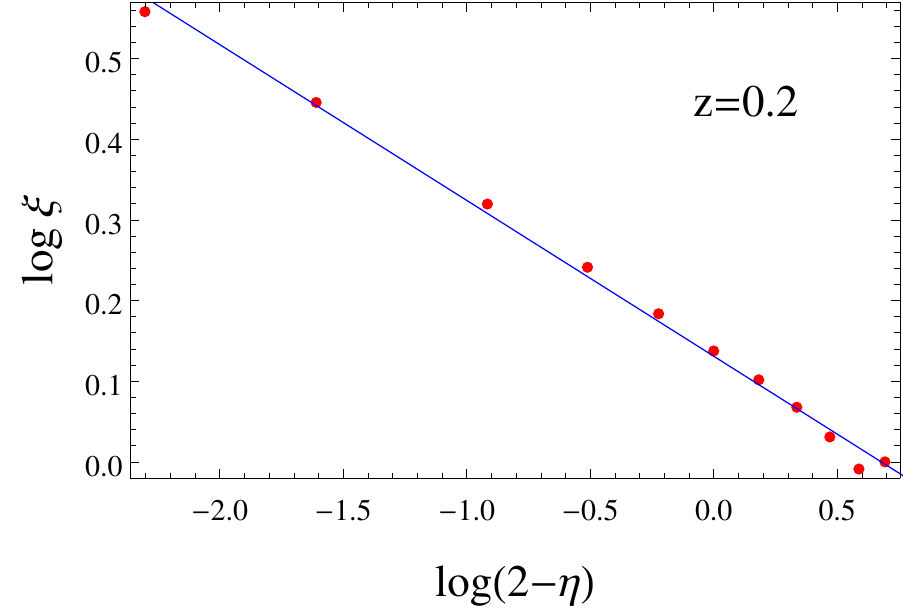}
	\caption{
	\textbf{Size of the texture $\xi$ around a single vacancy as a function of $2-\eta$ in the log-log scale.} (Color online) The size of the texture diverges as $(2-\eta)^{-z}$ with $z\approx 0.2$.
	}
	\label{fig:CorrLen}
\end{figure}

Since the zero moment condition of $\eta=2$ can be maintained everywhere only when there are four degenerate branches of states, for $0<\eta<2$, the picture changes. Within this range, the single vacancy ground states in a finite system are those with a $\psi_3$ background and a localized moment at the vacancy. The single vacancy ground states for different $\eta$ are shown in Fig.~\ref{fig:SV}. The panels illustrate that a length scale about the vacancy diverges as $\eta$ approaches $2$ with exponent $0.2$ (see Fig.~\ref{fig:CorrLen}) and that the texture becomes more anisotropic towards this highly degenerate point. This divergence reflects the simple fact that as one approaches the high symmetry point $\eta=2$ the line defects become less and less penalized energetically and extend further and further away from the vacancy.

We consider the cases of having two and three magnetic ions removed from a single tetrahedron in the pyrochlore structure. These two types of defect will contribute little for small random dilution and, for larger intermediate dilutions, the presence of single vacancy defects will dominate the behavior of the system so these remaining cases are mentioned largely for the sake of completeness. Suppose two magnetic ions are removed from a single tetrahedron. Evidently the remaining two sites have anti-aligned spins and by solving the zero moment conditions Eq.~\eqref{eqn:ZeroMoment} for the remaining two spins we note that these spins have equal angles lying within the set  $\theta=\pi/6 + n\pi/3$. On the lattice, a double vacancy on one tetrahedron implies the presence of two neighboring tetrahedra with single vacancies. So the ground states for a double vacancy have a $\psi_3$ background (up to line modes avoiding the vacancies) and a pair of line defects coming out of each of the tetrahedra with single vacancies.

Of course, when three sites are removed from a tetrahedron, the moment is necessarily non-zero on that tetrahedron. Then the minimum energy state will have some non-trivial spin texture. The form of this texture comes about as follows. When two sites are removed from the tetrahedron, we have seen that there are four line defects emerging from the vacancy sites. When a further spin is removed, the line defects remain except that there is a crossing point of the defect lines at the orphan spin (a lone spin on a tetrahedron) which leads to spin texture that is described in more detail in Section~\ref{ssec:texture}.

It is interesting to compare these results with other systems. In a conventional collinear antiferromagnet, one may remove sites with no effect on the neighboring spins since the direction of the local fields does not change. In a non-collinear antiferromagnet, removal of a single spin leads to a spin texture that decays away from the vacant site.\cite{villain1979insulating} Such a case is realized, for example, in the triangular lattice Heisenberg antiferromagnet.\cite{wollny2011fractional} At the opposite extreme, in the Coulomb phases in the Heisenberg model on a pyrochlore lattice, provided one leaves at least two occupied sites on a tetrahedron one does not introduce a net moment and the Coulomb phase remains even with a finite defect density. However, an orphan spin leads to a net moment and a non-trivial spin texture. The latter has been explored in kagome bilayers as in the material SCGO.\cite{sen2011fractional,sen2012vacancy} Just as the degeneracy $O(L^2)$ at $\eta=\pm 2$ in our case is intermediate between that of Coulomb phases and conventional magnets, so the effect of introducing vacancies lies in between the two extremes. 

We emphasize that we have considered only the case of vacancies in a finite system. We describe the approach to the thermodynamic limit and finite density in Section~\ref{sec:finite_density}. We shall find that results from this section provide useful insight into the finite density problem when $0<\vert \eta\vert<2$.

\subsection{Texture}
\label{ssec:texture}

Here, we consider $\eta=2$. We wish to understand how the dilute line case considered in the previous subsection evolves as the system size increases holding the vacancy concentration constant. As we do this, there will come a point at which two line defects with incompatible angles will intersect. In our model, this situation leads to the simplest non-trivial spin texture in which non-zero moments enter the system. In this section, we consider the nature of this spin texture. We shall find that the quasi-one dimensional physics observed in the previous subsection persists even in the case of a line crossing. We place vacancies in the finite system in such a way that two line defects of different angles intersect and we choose an initial spin configuration with a uniform $\psi_3$ background, four line defects and a randomly chosen angle at the intersection point. 

\begin{figure}[htpb]
	\begin{subfigure}[][]{
	\includegraphics[width=0.95\columnwidth]{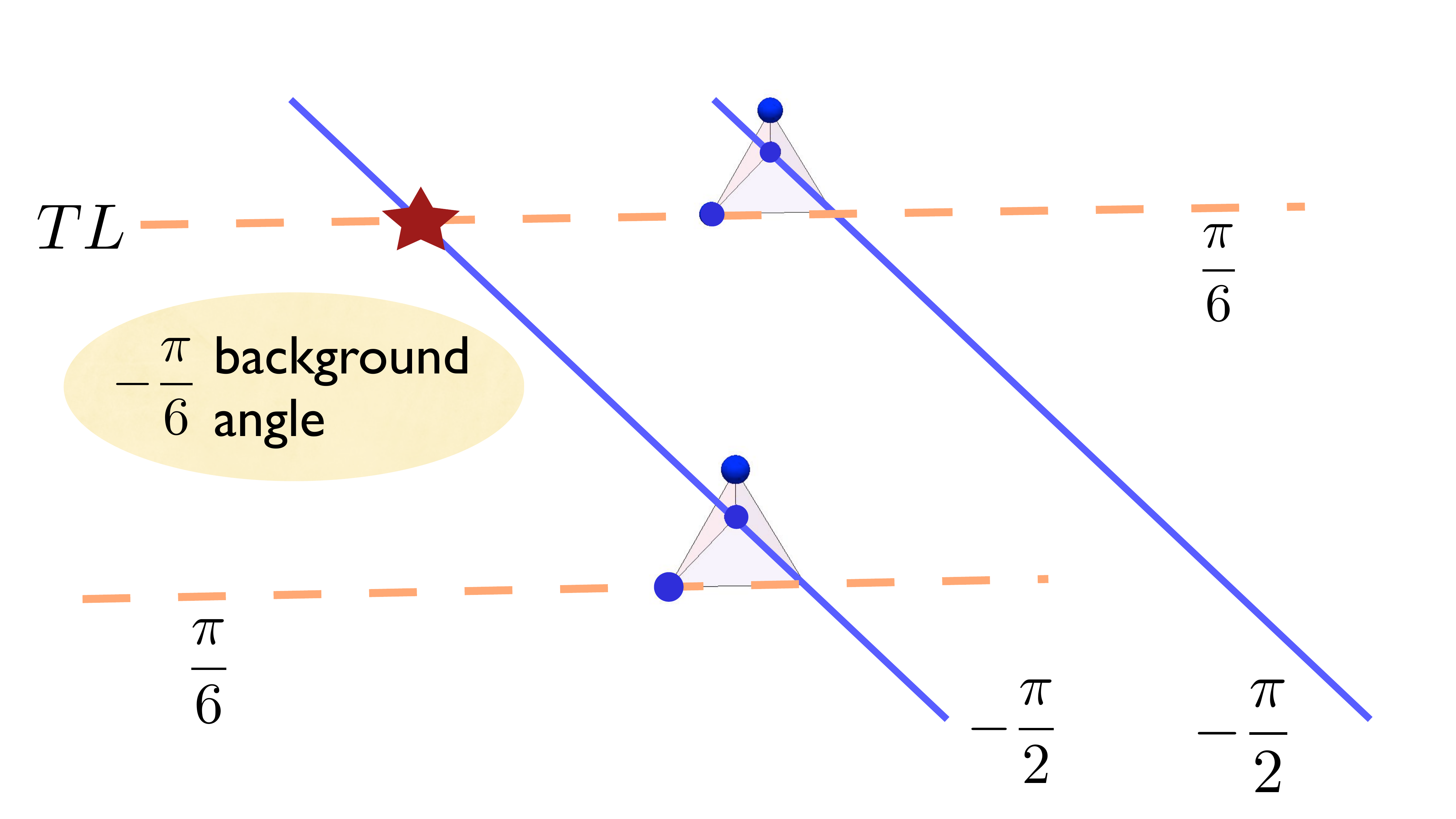}
}
	\end{subfigure}
	\begin{subfigure}[][]{
	\includegraphics[width=0.95\columnwidth]{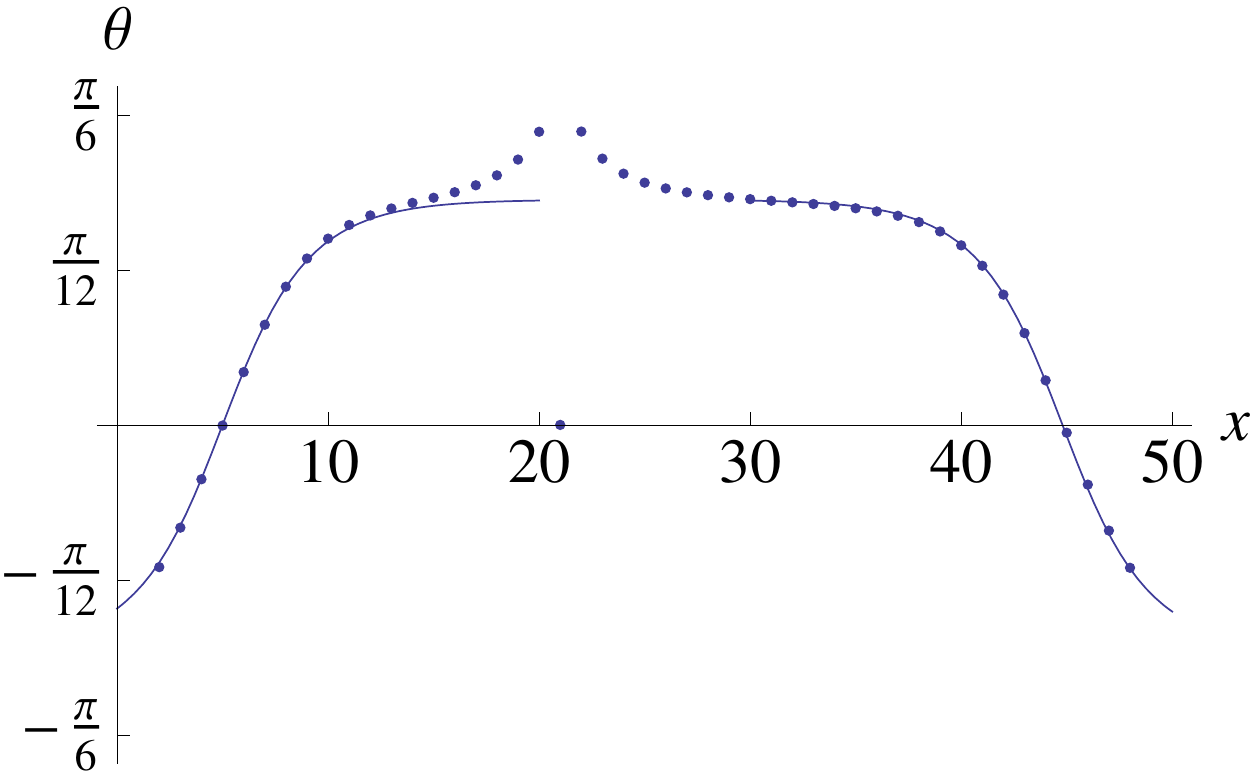}
}
	\end{subfigure}
	\caption{
	\textbf{ Simplest spin texture from intersecting line defects at $\eta=2$.} (Color online). 
	(a) Schematic plot of the spin texture. The figure shows a pair of pyrochlore tetrahedra each with three spins (blue circles) and with one missing spin. All other magnetic sites on the lattice are suppressed. Recall that a single vacancy has a ground state with a uniform background angle on all spins except for two line defects on which the angles of the spins in the local frame are different. Here, two line defects from different vacancies and with different local angles are brought to intersect. In this case, the minimum energy state has a uniform background angle with four line defects as illustrated in the figure. Three of these line defects have almost constant angle inherited from the single vacancy case. The fourth which we label $TL$ carries the non-trivial texture which is essentially one dimensional. All angles in this plot refer to the local coordinate frame and are the angles of the single vacancy case before minimization of the energy. Blue, solid lines carry the $\theta=-\pi/2$ angle and the orange, dashed lines carries $\theta=\pi/6$. The main text describes the effect on the angles.
	(b) Plot showing the angle along line $TL$. The horizontal axis $x$ is in units of the nearest neighbor spacing. The zero angle corresponds to the vacancy location. The peak appears in the vicinity of the intersection point. There are kinks on both sides because the system is periodic. The solid line belongs to a sine-Gordon kink based on background angles along defect line $TL$.
	}
	\label{fig:Texture}
\end{figure}

To be concrete, we take out two sublattice $1$ spins. We place the background angle in the $\psi_3$ configuration $-\pi/6$ and run two pairs of line defects through the system. One type of line defect carries angle $-\pi/2$ and the other $\pi/6$. We can choose the positions of the vacancies so that there is an intersection of $-\pi/2$ and $\pi/6$ lines at a single site. Figure~\ref{fig:Texture}(a) illustrates the nature of the initial state. From this starting configuration, we numerically minimize the energy by rotating spins into the local field.

The resulting spin texture has the following form. The background angle rotates into a new (almost) uniform angle. Four line defects remain, three of them almost unchanged except for an overall rotation in the local frame. All four defect lines are one-dimensional to an excellent approximation. One of the line defects containing the intersection point does change significantly - line $TL$ in Fig.~\ref{fig:Texture}(a). The choice of which one depends on the choice of initial state - in particular the initial angle at the intersection point: The two possibilities are degenerate. Figure~\ref{fig:Texture}(b) shows the texture along this line. The local fields at the intersection point and vacancy pin the spins at different angles. In between, the spin texture has a kink as it flips from one angle to another. There is a kink on both sides of the vacancy since the line is in a periodic system. Most of the tetrahedra along this line are in the $(\theta,\theta,-\theta,-\theta)$ branch leading to an approximately zero moment on tetrahedra along the line. Non-vanishing tetrahedral moments along this line are concentrated at the vacancy, the intersection point and at the kink centers.

In a non-periodic system, one of the line defects coming from infinity passes through the vacancy and then has a kink between the vacancy and intersection point. On the other side of the intersection point, the angle jumps to the background angle. In other words a semi-infinite segment of the line defect is healed by the intersection point.

Since the spin texture in our model is one-dimensional and since the exchange field coming from tetrahedral sites neighboring the line defect is uniform except at the vacancy and intersection points, and also because the model is ferromagnetic in the local coordinate frame, we expect the texture to be described by a sine-Gordon kink of the form $\arctan\exp\left( \alpha (x-x_0) \right)$ where $\alpha$ is a length scale related to the spin stiffness. This is because the XY ferromagnetic chain maps onto a sine-Gordon model in the continuum limit. Since the line defect angles maintain almost zero moment by belonging to tetrahedra in the $(\theta,\theta,-\theta,-\theta)$ branch of Eq.~\eqref{eq:eta2} with the uniform branch nearby, there is a cosine potential coming from the external exchange field with minima at the background angle and minus the background angle which accounts for the locations of the plateaux in the kink. A fit of the spin texture to the sine-Gordon kink is shown in Fig.~\ref{fig:Texture}(b).

\section{Finite Density}
\label{sec:finite_density}

In this section, we discuss the nature of the ground states in the thermodynamic limit. Over the range $0<\eta<2$, using the single vacancy problem as a guide, we expect the ground states to exhibit $\psi_3$ order. This is because the introduction of a single vacancy into a finite system breaks the continuous degeneracies leading to a $\psi_3$ background configuration. In the vicinity of each vacancy, there is a non-zero tetrahedral moment density that decays into the background away from the vacancy with some length scale. Then, because a single vacancy controls the nature of the background angle, over the range $0<\eta<2$, the single vacancy problem is expected to reflect the behavior of the model at finite density in the thermodynamic limit. In other words, there should be a $\psi_3$ background over this range at least for sufficiently dilute defects.

At the point $\eta=2$, a single vacancy in a finite system leads to a $\psi_3$ background configuration but with line defects instead of finite spin textures. In this case, the approach to the thermodynamic limit is more subtle. For $\eta=2$, a naive view of the approach to the thermodynamic limit holding the density of vacancies fixed has every chain ending on a vacancy leading to a configuration in which lines are dense. The background at finite density in an infinite system, if any, need not bear any relation to the single vacancy case. In this naive view, we begin with a system with dilute line defects and add a vacancy. Suppose the density of spins at the background angle is $f\equiv N_\mathrm{bg}/N$ where $N$ is the total number of sites. Adding a new vacancy seeds new defect lines, so $\Delta f=-\frac{L}{N}\left(\frac{N_\mathrm{bg}}{N}\right)$ implying that $f\sim \exp(-nL)$ where $n$ is the density of vacancies. Then the lines are dense when $nL\gtrsim 1$ - far from the thermodynamic limit. In fact, empirically, the number of background sites with no defect lines running through them is negligible for $nL\sim 1$. This observation makes quantitative the subtlety of studying the thermodynamic limit but also offers a resolution to the problem of  how to study the ground states numerically in a finite system. Parameter $nL$ controls the crossover between the dilute and dense line limits. We have found that choices of clusters and dilutions such that $nL\sim 1$ are sufficient to have quantities that scale with dilution $n$ only.

In the following, we present numerical and analytical evidence to confirm that $\psi_3$ selection indeed occurs over the range $0< \eta\leq 2$. From the symmetry about the locally isotropic point $\eta=0$, it would follow that $\psi_2$ is selected by vacancies over the range $-2 \leq \eta < 0$ in the ground state. Once again this state is ``opposite" to the state - $\psi_3$ - which is selected by thermal fluctuations over this range.

\subsection{Numerical Results for $\eta=2$}

In order to find ground state candidates, we employed the usual method of starting with some initial configuration and then iteratively aligning spins with their local fields:\cite{walker1980computer}
\begin{gather}
	\vec{\mss}_i = \vec{\mathsf{H}}_i/|\vec{\mathsf{H}}_i|\\
	\vec{\mathsf{H}}_{i\alpha} = \sum_{j\beta} J_{ij}^{\alpha\beta}\vec{\mss}_{j\beta}
\end{gather}
The iterations were performed until some convergence criterion was met. We stopped the iterations after the largest update, i.e. $\max_i||\vec{\mss}_\text{i,new} - \vec{\mss}_\text{i,old}||$ was smaller than $10^{-8}$. An important issue that affects the quality of the minimization is the choice of the initial states to be minimized. Due to frustration there is a large quasi-degeneracy and the final states after the minimization are dependent on the initial configurations. We used several types of initial states and picked the final state with the lowest energy and averaged results over distinct random disorder realizations. We found that the choice of the initial states that gives the lowest energy after the minimization depends on the value of $\eta$. 

Close to the $\eta=2$ point, the best initial states are uniform $\psi_3$ states with vacancies and sometimes $\psi_2$ states. For $\eta=2$, states with line defects should also be included for small $nL$, as suggested by the few vacancy study (Section~\ref{sec:VandT}). For larger $nL$, the contribution of these states to the disorder average falls off dramatically.\footnote{A point particularly relevant for $nL<1$ is that the angles at the crossings of line defects in the initial states can be fixed in different ways which also influences our choice of initial state.} It is worth pointing out that at $\eta=2$, the large low energy quasi-degeneracy means that a good guess for the initial state is crucial to find good ground state candidates: for example we have found consistently that the minimization procedure does not find the true ground state for a system with a single vacancy for sizes bigger than $L>3$ when starting from a uniform or a random state. That being said, we have averaged every system size and dilution over $100$ disorder realizations. The ground state energy scales linearly with dilution $n$. Also, a check of states with energies higher than our best ground state candidates suggest that the results that we find below seem to be shared by states of slightly higher energy. In all of this numerical work, we use cubic clusters with periodic boundary conditions, $16$ sites per cubic unit cell and edge length $L$ meaning $16\times L^3$ spins in total. We studied system sizes between $L=3$ and $L=12$. 

The considerations above suggest that, in order to approach the thermodynamic limit, we should concentrate our study on cluster sizes $L$ and dilutions $n$ with $nL\gtrsim 1$.  For clusters with $nL\leq 1$, candidate ground state configurations consist of line defects with crossings. As $nL$ increases, there is a sharp drop in the fraction of ground states obtained from initial states prepared by seeding line defects from each vacancy. For $nL\sim 1$ the lowest energy states we found tend to arise from a $\psi_3$ initial state. 

In the following, we consider the order parameter for  $\psi_2$ and $\psi_3$. We look at two quantities: the sublattice magnetization $M$ and the XY order parameter $O_{\psi_{2/3}}$
\begin{gather}
	m_a = \frac{1}{N_{a\text{s}}}\sqrt{\left(\sum_{p=1}\nolimits'\vec{\mss}_{pa}\right)^2}\notag\\
	\label{eqn:Msl}
	M = \frac{1}{4}\sum_{a=1}^4 m_a \\
	O_{\psi_{2/3}} = \frac{1}{N_\text{s}}\sum_{pa}\nolimits' \cos(6\theta_{pa}),
\end{gather}
where the prime implies the summation over all non-vacant sites; $N_\text{s}$ and  $N_{a\text{s}}$ are the total number of spins and the number of spins present in sublattice $a$ of the pyrochlore lattice respectively. The sum over $p$ runs over all fcc lattice vectors. When $O_{\psi_{2/3}}<0$, the angles are those belonging to the set of $\psi_3$ states and, for  $O_{\psi_{2/3}}>0$, the angles belong to the set of $\psi_2$ states. When both $O_{\psi_{2/3}}$ and $M$ are non-zero, $\psi_2$ or $\psi_3$ are selected.

\begin{figure}[htpb]
	\includegraphics[width=0.95\columnwidth]{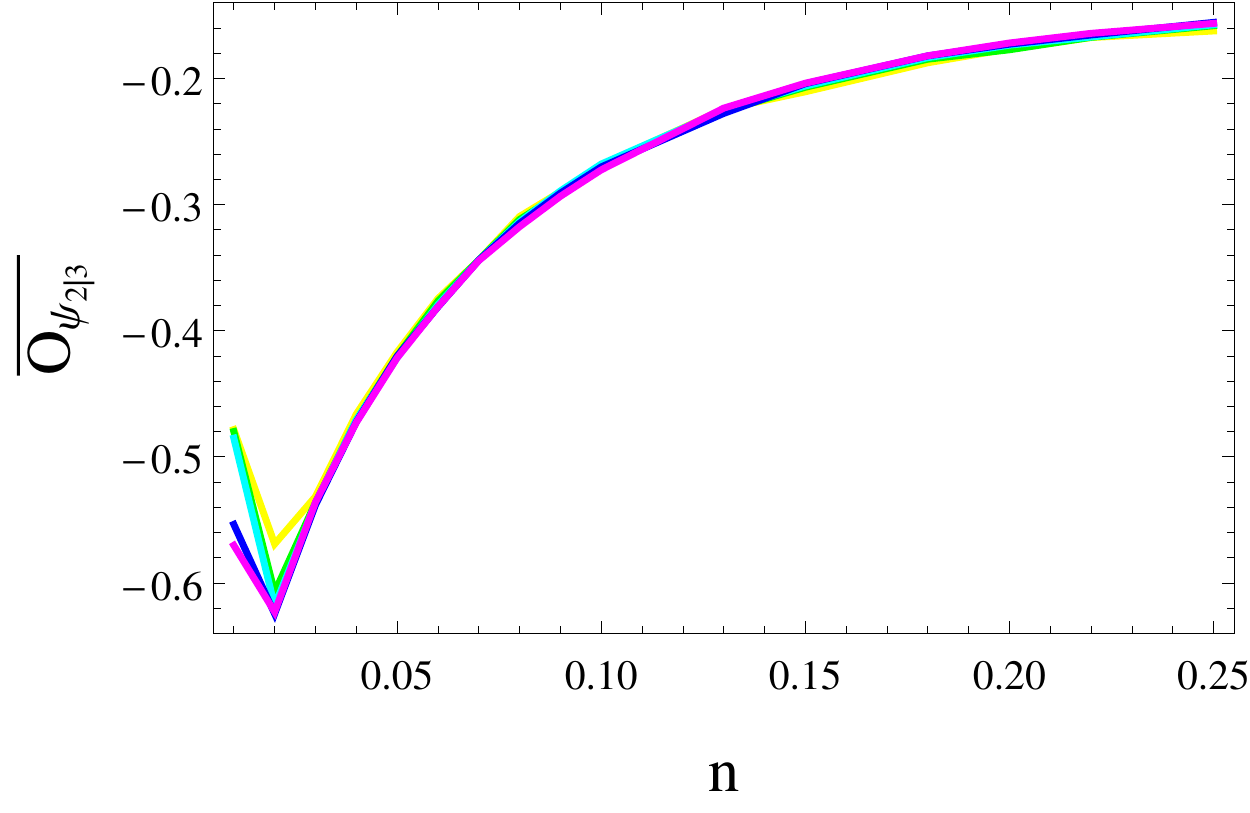}
	\caption{
	\textbf{Order parameter for $\psi_3$ of the disorder averaged candidate ground states $\eta=2$ for various dilutions.} (Color online). The different curves correspond to different system sizes $L=8-12$ corresponding to between $8192$ and $27648$ spins in the undiluted system. The overbar denotes the disorder average.
	}
	\label{fig:OP_vs_n}
\end{figure}

Figure~\ref{fig:OP_vs_n} shows the $O_{\psi_{2/3}}$ order parameter for $\eta=2$ and varying dilutions with different curves corresponding to a range of system sizes. The value of the order parameter confirms the selection of $\psi_3$. For dilutions $n$ greater than $2$ percent, the curves for different system sizes lie roughly on top of one another whereas for the smaller dilutions this ceases to be the case. One can understand this observation on the basis of the $nL$ crossover between the dilute line regime and the dense line limit described at the beginning of Section~\ref{sec:finite_density}. In particular, for the smallest dilutions and for the given range of system sizes, $nL\sim 1/10$, which puts the system in the dilute line regime. In this case, we should not expect the order parameter to scale with $n$. For larger dilutions, the systems are within the dense line regime which more correctly reflects the properties of the ground states in the finite density thermodynamic limit. As one would expect, the value of the order parameter falls off as the dilution increases: The distribution of angles (see Fig.~\ref{fig:FD-Angle-Histogram}) gets broader with less pronounced peaks at $\psi_3$ angles.

So far we have studied the ground states at $\eta=2$ numerically but without providing an understanding of the approach to the infinite system limit at finite density from the single vacancy problem. Some insight can be gained from the spin texture that we studied in Section~\ref{ssec:texture}. We have seen that defect lines radiating from isolated vacancies must  cross as the system size is increased at finite vacancy density and an energetic cost is incurred at vacancies and the crossing points between these defect lines. The important observation here is that there is healing of line defects into the background angle when a line defect intersects with another with incompatible local angles. This hints at the possibility that $\psi_3$ selection, which occurs for a single vacancy in a finite system at $\eta=2$, is somewhat robust to line crossings and occurs also in the infinite system ground state of the $\eta=2$ system with a finite concentration of vacancies. 

\subsection{Generalized XY Model}

Away from $\eta=2$ point, in addition to uniform ($\psi_2$ and $\psi_3$) angle initial states, random initial states can also give the best ground state energy for certain disorder realizations. It appears that when some of the clean system degeneracy is removed by going away from $\eta=2$, our minimization procedure can more efficiently find the ground state. We used $500$ random initial states for each value of $\eta$ and dilution away from the high degeneracy point. We have performed minimization for all these different initial states and selected the final state having the lowest energy. As before, every system size and dilution was averaged over $100$ disorder realizations. 

\begin{figure}[htpb]
	\includegraphics[width=0.95\columnwidth]{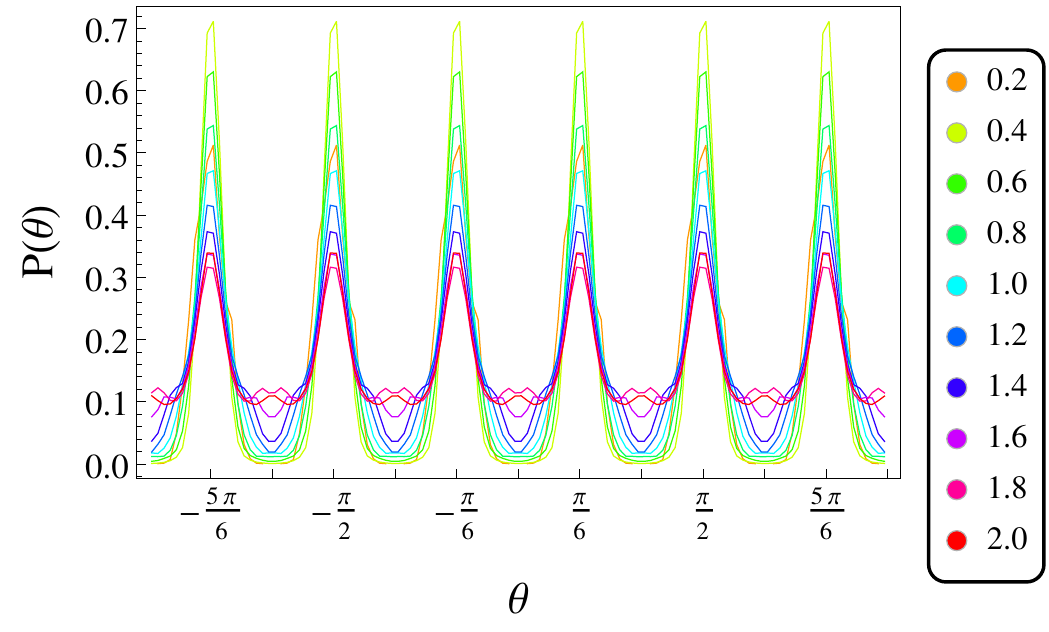}
	\caption{
	\textbf{Symmetrized histograms of angles for different values of $\eta$ (shown in the legend) for $L=12$ and a dilution of $8\%$.} (Color online).There are peaks at the $\psi_3$ angles $\pi/6 + n\pi/3$. As $\eta$ tends toward $\eta=2$, the peaks around the $\psi_3$ angles become less sharp reflecting the larger quasi-degeneracy of the system as more configurations enter the ground state. Also as $\eta\to0$ the peaks flatten out: at the isotropic point $\eta=0$ the histogram becomes flat. We see this decrease already for $\eta=0.2$, which has smaller peak heights as compared to the case $\eta=0.4$.
	}
	\label{fig:FD-Angle-Histogram}
\end{figure}

The disorder averaged histogram of angles in the local coordinate frame (see Fig.~\ref{fig:FD-Angle-Histogram}) has clear peaks at the angles $\pi/6+n\pi/3$ which are those associated with the $\psi_3$ configurations. In the ground state minimizations, we bias the $\psi_3$ initial state towards one of these angles. The resulting angle histogram retains this bias. Since the problem is completely symmetric under $\pi/3$ rotations, we symmetrize the resulting histogram, the result of which is shown in Fig.~\ref{fig:FD-Angle-Histogram}. The histograms clearly indicate $\psi_3$ selection. The $\psi_3$ peak height grows as $\eta$ decreases until $\eta\approx0.4$ beyond which the peaks start to decrease reflecting the absence of selection at the isotropic point $\eta=0$ at which point any uniform state is a ground state.

\begin{figure}[htpb]
	\includegraphics[width=0.95\columnwidth]{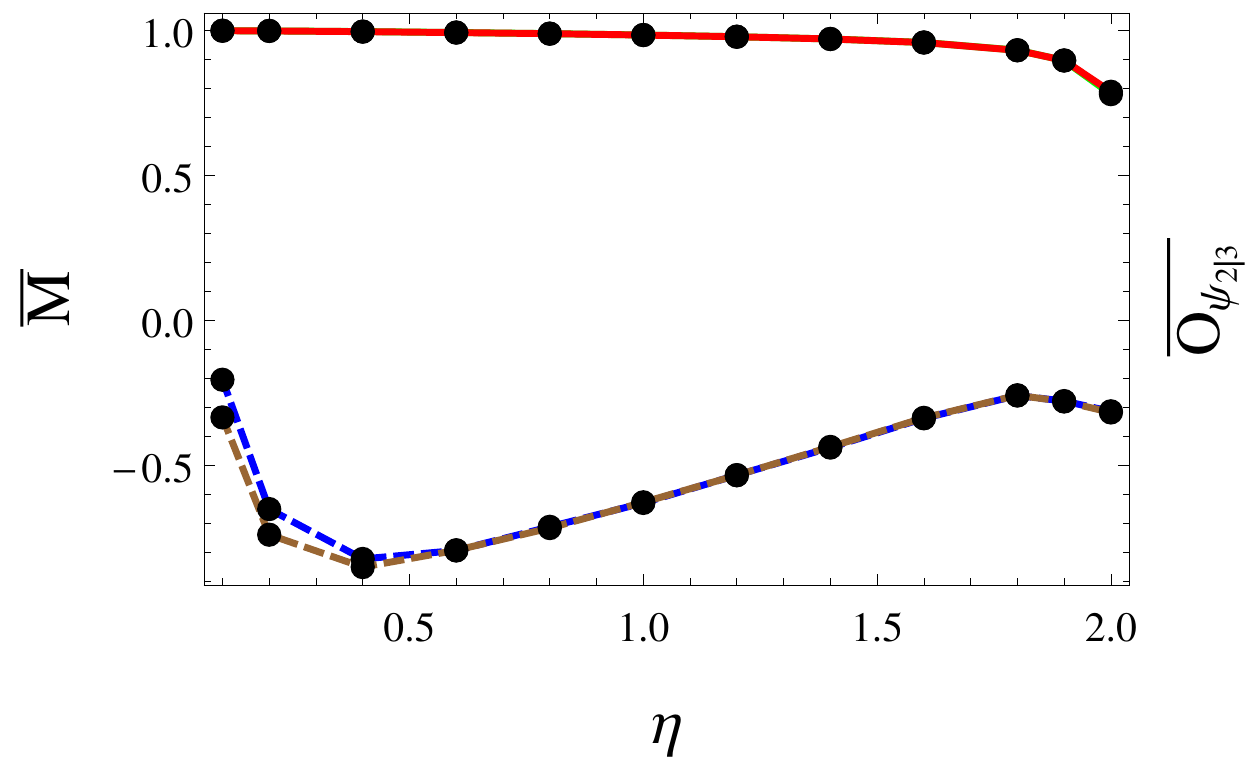}
	\caption{
	\textbf{Order parameter for $\psi_3$ of the disorder averaged candidate ground states for various $\eta$ with $L=10$ and $12$.} The blue and brown (bottom, dashed with points) curves corresponds to disorder averaged $\overline{O_{\psi_{2/3}}}$ and the red (upper, solid) curve is the disorder averaged sublattice magnetization, $\overline{M}$, as defined by~Eq.~\eqref{eqn:Msl}. The non-monotonic behavior close to $\eta=0$ is due to the absence of selection at $\eta=0$: any state with a uniform angle $\theta$ is a ground state.
	}
	\label{fig:OP_vs_eta}
\end{figure}

Figure~\ref{fig:OP_vs_eta} shows the order parameter of the $\psi_3$ state for different values of $\eta$ in the range $0 < \eta\leq 2$ for two different system sizes $L=10$ and $12$. The weight of numerical evidence is that $\psi_3$ state is selected by dilution over the entire range $0 < \eta\leq 2$.  This is the main result of this work. The order parameter varies non-monotonically with $\eta$ which reflects (i) the absence of dilution selection at $\eta=0$ which is a ferromagnet in the local frame and (ii) the anomalously high degeneracy in the clean case at $\eta=2$. 

\subsection{Perturbative Calculation}
\label{sec:PT}

\begin{figure}[htpb]
	\includegraphics[width=0.95\columnwidth]{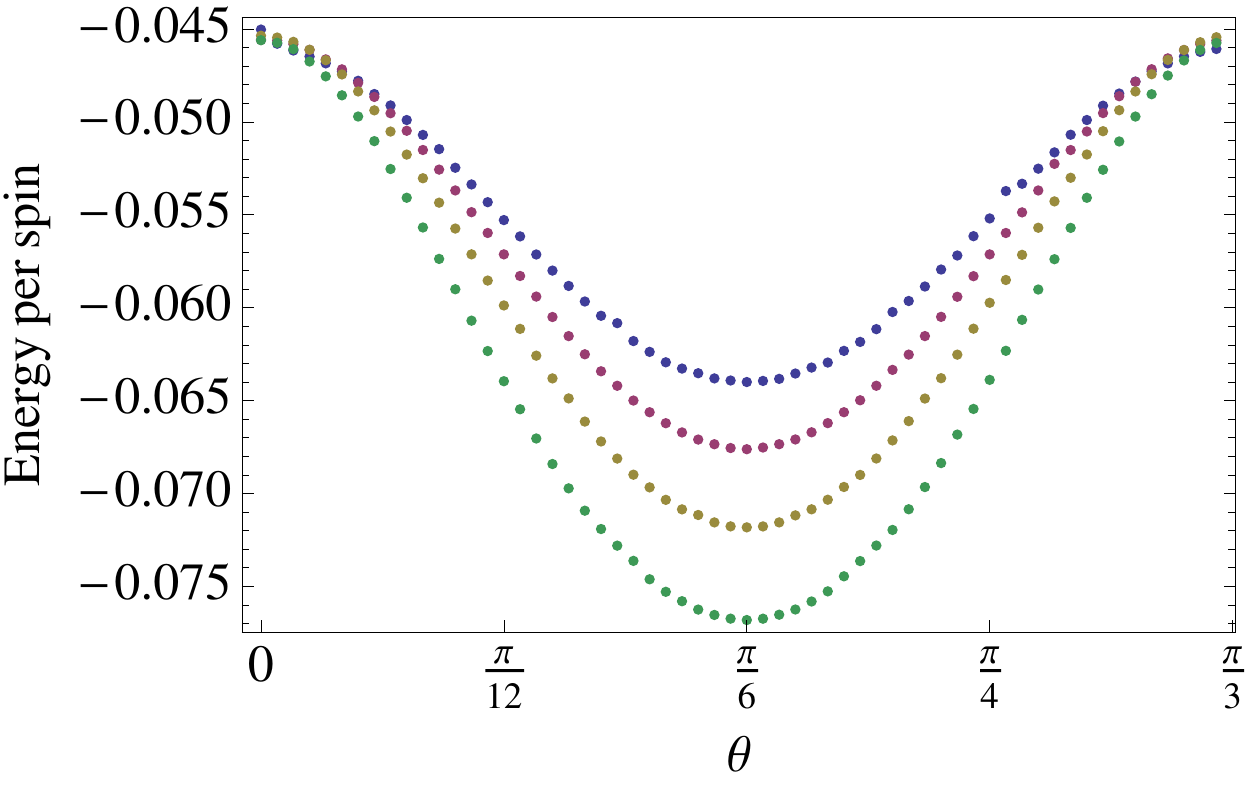}
	\caption{
	\textbf{ Results of the perturbation theory calculation described in the main text.} (Color online) The plot shows the shift in the ground state energy per spin relative to the clean system ground state due to the presence of a finite density of vacancies. Different curves correspond to different values of $\eta$. From bottom to top these are $1.6$, $1.4$, $1.2$, $1$.
	}
	\label{fig:PT}
\end{figure}

Here we present results of a perturbative calculation of the shift in the ground state energy induced by dilution. The idea is to approximate the problem of computing the ground states in the presence of vacancies to computing a quadratic energy shift in angular displacements. The quadratic approximation to the energy can be minimized exactly and disorder averaged. For the result we compute the energy shift due to dilution about the equal angle ground states of the clean system. This technique has been applied successfully in various cases.\cite{henley1989ordering,prakash1990ordering} The details of the perturbative calculation are given in Appendix~\ref{sec:appendix3} and the shift in the energy per spin is given by
\[
	\langle\delta\mth_\text{dil}\rangle  = -\frac{J^2}{2(1-n)^2}\sum_{ij}A_{ij}^{-1}(\theta)B_{ij}(\theta)
\]
where $A_{ij}$ is the Hessian matrix in real space defined in Section~
\ref{sec:ObD} and $B_{ij}$ depends on the disorder averaged bond-bond correlation function and the matrix of first derivatives of the Hamiltonian. The fraction of occupied sites is $1-n$. Figure~\ref{fig:PT} shows the energy shift per spin computed within this perturbative scheme in the case where $1-n$ is close to one - the dilution $n$ is small. The energy shift due to dilution has a minimum at the $\psi_3$ angles ($\pi/6$ in the figure) over the entire range $0<\eta \leq 2$. 

\subsection{Finite Temperature}
\label{sec:finite_temperature}

At finite temperature, thermal fluctuations for $0<\eta\leq 2$ lead to an ordering transition into the $\psi_2$ structure (see Fig.~\ref{fig:TcSketch}). As we have seen, dilution selects instead the $\psi_3$ ordered ground state structure. The competition between these mechanisms will lead to a phase transition between the two ordered phases in the classical model for small dilutions. In this section, we address the nature of the lower temperature transition originating from this competition. 

To do this, we consider, once again, a single vacancy in a finite system at $\eta=2$. At zero temperature, the configuration is one with a uniform $\psi_3$ background state modulo line modes with two line defects radiating from the vacancy. At non-zero temperature there is a finite size transition into the $\psi_2$ state with some spin texture around the vacancy. Using Monte Carlo simulations, we study the tetrahedron moment density around the single vacancy. By heating the system from the ground state, we find that the line defects disappear abruptly at some temperature which goes to zero as the system size increases as $1/L^3$ since the single vacancy costs only an energy of $O(J)$. Above this temperature, within the $\psi_2$ ordered state, the spin texture is strongly localized around the vacancy within two lattice spacings up to the finite size transition temperature. In summary, there is no apparent varying length scale around the vacancy within the $\psi_2$ phase: the transition is first order.

To understand this result, we consider the free energy supposing that there is some length scale, $\xi$, around the vacancy. In this case, there should be an entropic cost when a texture is present going like $T\xi^3$. The energetic cost should be computable from $\int d^3x (\nabla\phi)^2 \sim \xi$. Then the free energy has its minimum when $\xi=0$. On this basis, we expect the transition from $\psi_2$ to $\psi_3$ to be first order and since this argument does not depend on $\eta$, this result should hold over the entire $(T,\eta)$ phase diagram.

\section{Summary and Conclusions}
\label{sec:conclusions}

We have investigated the effects of dilution in an XY model on the pyrochlore lattice. At finite temperature, this model is known to exhibit order-by-disorder meaning that thermal fluctuations lift the classical degeneracies leading to discrete symmetry breaking at low temperatures. Quantum fluctuations have a similar effect. This physics is controlled by a pair of special couplings $\eta=\pm 2$ in the model at which the ground state spin configurations have a quasi one-dimensional nature exhibiting an $O(L^2)$ entropy. The phase diagram of the model has a symmetry that allows us to map results from $\eta=2$ to $\eta=-2$ by rotating the spins in their local frames through $\pi/2$ so that results for $0\leq \eta\leq 2$ map to $-2\leq\eta\leq0$ under a simple transformation.

Research into the effects of introducing point-like disorder into frustrated magnets has a long history (see, for example, \onlinecite{villain1979insulating,sen2011fractional,wollny2011fractional}). We observe that the introduction of single or double vacancies leads to line defects running through the system at the highly degenerate points of our model. Crossing points of these defects leads to a non-trivial spin texture which remains one-dimensional. 

We have tackled the problem of finding the ground states in infinite systems with a finite concentration of vacancies. We find that dilution alone selects ground states exhibiting long-range order that is different to that selected by thermal and quantum fluctuations. This conclusion can be reached over by studying the few vacancy problem of Section~\ref{sec:VandT}. However, we have also carried out numerical minimization of the energy and complemented this with an analytical study both directly at finite density.

Since there is competition between thermal and dilution-induced order-by-disorder, the phase diagram of the classical model should have two phase transitions as a function of temperature for small dilutions. The higher temperature thermal-fluctuation induced transition and a lower temperature transition from one ordered state to another. The former transition has been studied in Ref.~\onlinecite{zhitomirsky2014finite} showing that the first order transition at $\eta=2$ gives way to a continuous transition in the XY universality class with IR physics strongly influenced by the dangerously irrelevant $\psi^6$ operator in the Landau free energy. This is not expected to change for small dilutions. We argue in the main text that the lower temperature transition should be first order over the entire range $-2\leq\eta\leq 2$.

Our findings may have implications for materials among the rare-earth pyrochlores for example in Er$_2$Ti$_2$O$_7$ which exhibits the one known case of $\psi_2$ selection through order-by-disorder. In order to investigate the presence of physics describe in this work, one would dope out some proportion of the magnetic ions with non-magnetic impurities, for example by looking at Er$_{2-x}$Y$_x$Ti$_2$O$_7$ for a small concentration $x$ of yttrium,\cite{niven2014} to see whether it has a transition into $\psi_3$ below the $\psi_2$ transition. 

The magnetic exchange interactions in the material \eto\ have been very well characterized by fitting the spin wave excitations in a $3$T field along different directions and in zero field. The parameters were found to be
\begin{align*} 
	& J^{\pm\pm}=4.2\pm 0.5 \hspace{0.5cm}  & J^\pm=6.5\pm 0.75 \\
	& J^\mathrm{zz}=-2.5\pm 1.8 \hspace{0.5cm} & J^{z\pm}=-0.88\pm 1.5
\end{align*}
in units of $10^{-2} \mathrm{meV}$.\cite{savary2012order,oitmaa2013phase} The classical ground state for this model is branch $B1$ - the equal angle branch of Eq.~\eqref{eqn:GS1} - which is lifted by fluctuations.\cite{mcclarty2014order,wong2013ground,yan2013living} In the absence of the out of plane couplings, \eto\ has $\eta\approx 1.3$. We expect that the physics discussed in this paper is robust to the presence of the out-of-plane couplings. We believe it likely that dilution selects $\psi_3$ when long-range dipolar interactions are included because these interactions preserve the classical $U(1)$ symmetry of the equal angle branch $B1$ in the clean system\cite{savary2012order} and the soft modes for some range of couplings, both of which are central to the selection of long-range order by dilution and because diluted tetrahedra tend to minimize their moments. However, the problem of disorder selected long-range order in the presence of long-range interactions is one that deserves further investigation. 

Physical realizations of XY order-by-disorder occur in effective spin one-half models since the anisotropy is typically generated through the combination of spin-orbit coupling and crystal field leading to an anisotropic doublet. Quantum order-by-disorder in the XY pyrochlore model we have considered selects the same state as quantum fluctuations which therefore competes with the ground states selected by dilution. An important problem that remains to be tackled is to establish the nature of the disorder-driven phase diagram in the presence of quantum fluctuations. 

\bigskip

As we were concluding this work, a preprint appeared by V. Maryasin and M. Zhitomirsky\cite{maryasinzhitomirsky} that reaches similar conclusions on the selection of an ordered state by dilution.
 
\begin{acknowledgments}
	We would like to thank John Chalker for many useful discussions during the course of the investigation and both John Chalker and Michel Gingras for their comments on the manuscript. P.M. acknowledges a Keeley-Rutherford fellowship from STFC in association with Wadham College, Oxford. P.M. also acknowledges the programme "Novel Directions in Frustrated and Critical Magnetism" at NORDITA during which a part of this work was conducted.
\end{acknowledgments}

\appendix

\section{Local Coordinate Frame}
\label{sec:appendix}

The pyrochlore lattice can be viewed as an fcc lattice with a tetrahedral basis. We take the following convention for the (right-handed) local coordinate frame. The local $\mhz_a$ axes for sublattice $a$ are, in the lab frame
\begin{align*}
	\mhz_1 & = \frac{1}{\sqrt{3}} \left( 1,1,1 \right) \\
	\mhz_2 & = \frac{1}{\sqrt{3}} \left( 1,-1,-1 \right) \\
	\mhz_3 & = \frac{1}{\sqrt{3}} \left( -1,1,-1 \right) \\
	\mhz_4 & = \frac{1}{\sqrt{3}} \left( -1,-1,1 \right) 
\end{align*} 
and the  local $\mhx_a$ axes are 
\begin{align*}
	\mhx_1 & = \frac{1}{\sqrt{6}} \left( -2,1,1 \right) \\
	\mhx_2  & = \frac{1}{\sqrt{6}} \left( -2,-1,-1 \right) \\
	\mhx_3 & = \frac{1}{\sqrt{6}} \left( 2,1,-1 \right) \\
	\mhx_4  & = \frac{1}{\sqrt{6}} \left( 2,-1,1 \right).
\end{align*} 

In this frame the Hamiltonian is of the form given in the main text Eqn.~\eqref{eqn:H} with $\gamma$ matrix
\begin{gather}
	\gamma = \left(
	\begin{array}{cccc}
		0 & 1 & e^{2i\pi/3} & e^{4i\pi/3} \\
	    0 & 1 & e^{4i\pi/3} & e^{2i\pi/3} \\
		e^{2i\pi/3} & e^{4i\pi/3} & 0 & 1 \\
		e^{4i\pi/3} & e^{2i\pi/3} & 1 & 0
	\end{array}
	\right).
\end{gather}
The matrix $\zeta = -\gamma^{*}$.

\section{Single Vacancy for $\eta=2$}
\label{sec:appendix2}

Let us remove a spin on a single tetrahedron on site $1$ without loss of generality. The zero moment conditions are
\begin{align*}
	\cos\theta_2 - \cos\theta_3 - \cos\theta_4 = 0\\
	\cos \bar\theta_2  + \cos\bar\theta_3 - \cos\bar\theta_4  = 0\\
	\cos \hat\theta_2   - \cos\hat\theta_3   + \cos \hat\theta_4  = 0.
\end{align*} 
There is a discrete set of solutions to these equations: 
\begin{align*}
	\left( - , \pi/2, 5\pi/6, \pi/6  \right) \\
	\left( - , \pi/6,-3\pi/2, -\pi/6  \right) \\
	\left( - , -\pi/2, \pi/6, 5\pi/6  \right) \\
	\left( - , -\pi/6, \pi/6, \pi/2  \right), 
\end{align*}
and also those related to these under time reversal. We now turn to the lattice problem with a single vacancy which has two tetrahedra with a missing site and we recall the degenerate branches for the lattice without vacancies. 
\begin{align*}
	B1: & \left( \theta,\theta,\theta,\theta \right) \\
	B2: & \left( \theta,\theta,-\theta,-\theta \right)  \\
	B3: & \left( \theta,2\pi/3 - \theta,2\pi/3 - \theta,\theta \right)  \\
	B4: & \left( \theta,4\pi/3- \theta,\theta,4\pi/3 - \theta \right).
\end{align*}
Suppose we place the magnetic ions on the lattice into the equal angle branch with $\theta=-\pi/6$ and remove a spin on site $1$. Then the tetrahedra with vacancies have configurations $\left(-,\pi/6,-\pi/2,-\pi/6\right)$. Then we can pass a chain with alternating sublattices $1$ and $2$ through the vacancy and, along that chain the tetrahedra are locked into the configuration $(\pi/6,\pi/6,-\pi/6,-\pi/6)$ belonging to branch $2$. We also pass a chain with alternating sublattices $1$ and $3$ through the vacancy and the tetrahedra along that chain are in configuration $(-\pi/2,-\pi/6,-\pi/2,-\pi/6)$ belonging to branch $4$. All tetrahedra in the lattice have zero moment.
 
In general, the presence of a single vacancy in a finite system, forces the background angle to one of the $\psi_3$ configurations with two chains of tetrahedra in different branches. 

\section{Details of the Perturbation Theory with Dilution}
\label{sec:appendix3}

Suppose the spin interactions between nearest-neighbors for the general XY model take the form
\begin{gather}
	\mth = \sum_{\langle i,j \rangle} J_{ij}^{\alpha\beta} c_i c_j \mss_{i\alpha}\mss_{j\beta}
\end{gather}
where ${c_i}=0,1$ account for the effect of dilution and the moments ${\mss_i}$ are parametrized by angle $\theta_i$. Let $\langle c_i\rangle=p=1-n$ be the probability that a site is occupied so that $p=1$ corresponds to the clean system. Our approach to studying the effects of dilution is to expand in small angular deformations $\theta_i \rightarrow \theta_i +\delta\theta_i$  about a set of ground states dictated by the clean system: the equal angle branch for general $\eta$. Once we have expanded to quadratic order we can solve for the angular deviations in the diluted system and compute the total energy to this order.

To this end, we introduce $\Gamma_{ij} \equiv c_i c_j-p^2 $ which has a vanishing disorder average. In terms of this variable, the Hamiltonian is 
\begin{gather}
	\label{eqn:Hrw}
	\mth_p = p^2\mth +  \sum_{\langle i,j \rangle} J_{ij}^{\alpha\beta} \Gamma_{ij} \mss_{i\alpha}\mss_{j\beta}.
\end{gather}
Expanding in small displacements $p^2 H$ gives
\begin{gather}
	\delta\mth^{\left(1\right)}=p^2\frac{J}{2}\sum_{ij}\delta\theta_i A_{ij}\delta\theta_j
\end{gather}
where we have pulled out an overall scale and $A_{ij}$ is the Hessian. The second term on the right-hand-side of Eq.~\eqref{eqn:Hrw} gives
\begin{gather}
	\delta\mth^{\left(2\right)}=\sum_i\gamma_i\delta\theta_i
\end{gather}
where $\gamma_i$ is related to the matrix of first derivatives of the Hamiltonian $X_{ij}$. 

Minimizing with respect to the $\delta\theta_i$ we find
\begin{align}
	\delta\theta_i & = \frac{J}{p^2} \sum_j A_{ij}^{-1}(\theta)F_j(\theta) \\
	\label{eqn:delta_theta}
	F_j & = \sum_k \eta_{jk} X_{jk}
\end{align}
and substituting back into the Hamiltonian gives 
\begin{align*}
	& \langle\delta\mth_\text{dil}\rangle =\\
	& =-\frac{J^2}{2p^2}\sum_{ij}A_{ij}^{-1}(\theta)\sum_{k\in\partial i}\sum_{l\in\partial j}\langle\Gamma_{ik}\Gamma_{jl}\rangle X_{ik}(\theta)X_{jl}(\theta)\\
	& = -\frac{J^2}{2p^2}\sum_{ij}A_{ij}^{-1}(\theta)B_{ij}(\theta)
\end{align*}
where $B_{ij}\equiv p^2(1-p^2)B^{(1)}_{ij} + p^3(1-p)B^{(2)}_{ij}$. For small dilutions $x=1-p$, we have $B_{ij}\equiv 2x B^{(1)}_{ij} +  x B^{(2)}_{ij}$. The disorder average gives us
\begin{gather}
	\langle\Gamma_{ik}\Gamma_{jl}\rangle=
	\begin{cases}
		p^2(1-p^2) & ik\,\text{and}\: jl\,\text{are the same bond}\\
		p^3(1-p) & ik\,\text{and}\: jl\,\text{have one site in common}\\
		0 & i\neq j,l\,\text{and}\, k\neq j,l
	\end{cases}
\end{gather}

Written explicitly we have
\begin{align*}
	B^{(1)}_{ij} & = X_{ij}(\phi)X_{ji}(\phi) + \delta_{ij}\sum_{k\in\partial i}X_{ik}(\phi)X_{ik}(\phi) \\
	B^{(2)}_{ij} & = \delta_{ij}\delta_{ab}\sum_{l,k\in\partial i}\sum_{l\neq k}X_{ik}^{ad}(\phi)X_{il}^{ac}(\phi) \\
	& +\sum_{j,k\in\partial i}\sum_{j\neq k}X_{ik}^{ac}(\phi)X_{ji}^{ba}(\phi) + \sum_{i,l\in\partial j}\sum_{l\neq i}X_{ij}^{ab}(\phi)X_{jl}^{bd}(\phi) \\ 
	& +\sum_{k}\sum_{i,j\in\partial k}\sum_{i\neq j}X_{ik}^{ac}(\phi)X_{jk}^{bc}(\phi)
\end{align*}
where $\partial i$ denotes the set of nearest-neighbors of site $i$.

We Fourier transform and evaluate the resulting reciprocal space sum leading to the results given in Section~\ref{sec:PT} and especially Fig.~\ref{fig:PT} which shows the leading order result for small $1-p$.

\bibliography{Dilution}

\end{document}